\documentclass[pra,aps,twocolumn,showpacs]{revtex4}
\usepackage{bbm,amssymb,amsmath,graphicx,color,times}

\newcommand{\ket}[1]{\vert #1 \rangle}
\newcommand{\ketbra}[2]{\vert #1 \rangle \langle #2 \vert}
\newcommand{\braket}[2]{\langle #1  \vert #2 \rangle}

\bibstyle{prsty}

\newcommand*\xbar[1]{\hbox{\vbox{\hrule height 0.3pt \kern0.35ex 
\hbox{\kern-0.0em \ensuremath{#1}\kern-0.0em}}}} 
\begin{document}
\title{Quantum metrology in Lipkin-Meshkov-Glick critical systems}
\author{Giulio Salvatori}
\author{Antonio Mandarino}
\affiliation{Dipartimento di Fisica, Universit{\`a} degli 
Studi di Milano, I-20133 Milan, Italy}
\author{Matteo G.A. Paris}\email{matteo.paris@fisica.unimi.it}
\affiliation{Dipartimento di Fisica, Universit{\`a} degli Studi 
di Milano, I-20133 Milan, Italy}
\affiliation{CNISM, Udr Milano, I-20133 Milan, Italy}
\begin{abstract}
The Lipkin-Meshkov-Glick (LMG) model describes critical systems 
with interaction beyond the first-neighbor approximation. Here 
we address quantum metrology in LMG systems and show how 
criticality may be exploited to improve  precision. 
At first we focus on the characterization of LMG systems themselves, 
i.e. the estimation  of anisotropy, and address the problem by 
considering the quantum Cram\'er-Rao bound. We evaluate the 
Quantum Fisher Information of small-size LMG chains made of $N=2, 3$ 
and $4$ lattice sites and also analyze the same quantity in the 
thermodynamical limit. Our results show that criticality is indeed
a resource and that the ultimate bounds to 
precision may be achieved by tuning the external field and  
measuring the total magnetization of the system. We then address the 
use of LMG systems as quantum thermometers and show that: i) precision 
is governed by the gap between the lowest energy levels of the systems, 
ii) field-dependent level crossing is a metrological resource to extend 
the operating range of the quantum thermometer. 
\end{abstract}
\maketitle
\section {Introduction}
During the last decade a plentiful contamination between condensed
matter physics and quantum information theory has been exploited. 
On the one hand, many body systems exhibiting quantum phase 
transitions (QPTs), usually studied in terms of order parameters, 
correlation lengths and symmetry breaking~\cite{Sachdev} have been 
fruitully analyzed in terms of quantum information based tools, such 
as dynamics of correlation in the ground state (GS) of the 
systems~\cite{EntSpin} and quantum information geometry 
\cite{Zan1,Zan2,Zan3,Zan4,FisherOrder}. On the other hand, 
quantum critical systems have been shown to provide a resource 
for quantum estimation and metrology, offering superextensive precision in the
characterization of coupling parameters and thermometry \cite{ZP,Inv1,Inv2}.
\par
The keystone of quantum estimation theory (QET) resides in the quantum
version of the Fisher Information~\cite{Helst,LQE}, a measure that
accounts for the statistical distinguishability of a quantum state from
its neighboring ones. Indeed, the geometrical approach to QPT has shown
how to improve estimation strategies for experimental inaccessible
parameter by driving the system towards critical points, where a sudden
change in the ground state structure takes place~\cite{Zanardi, ZP}. In
particular this behaviour has been tested in models where the
interaction is restricted to first neighbors~\cite{Inv1,
Inv2, XYdis}, e.g. quantum Ising and X-Y models in an external 
field, in order to precisely estimate the parameters of the system 
and to provide useful information about the phase diagram.
In view of the attention paid to systems with more sophisticated
interaction among lattice sites \cite{NCl, Nori, Nmag, Zan5} a question
thus naturally arises on whether criticality may be exploited to enhance
metrology in systems with interaction beyond the first-neighbor
approximation. 
\par
In this framework, 
systems described by the Lipkin-Meshkov-Glick (LMG) model provide non
trivial examples to assess quantum criticality as a resource for quantum
estimation. LMG was first proposed as a simple test for many-body
problems approximations~\cite{LMG1,LMG2,LMG3} and since then it has been
used to describe the magnetic properties of several molecules,
remarkably Mn$_{12}$Ac~\cite{Mn12}.  It also found applications in
several different fields, leading to a variety
of results in terms of entanglement properties of its ground
state~\cite{vid1, vid2, vid3} and spin squeezing~\cite{SpinSqLMG}.
For finite size chains LMG have been characterized in terms  of fidelity
susceptibility~\cite{FidSusLMG1, FidSusLMG2,cas12} and adiabatic
dynamics~\cite{can08,can09,sol08}. Although the LMG model cannot be
solved analytically for a generic chain size, some of its extensions are
amenable to an exact solution \cite{fue07}. We also mention that the LMG 
model received attention not only theoretically: experimental 
implementations have been proposed using condensate systems in a double 
well potential~\cite{BoseEinsteinDouble} or in
cavities~\cite{BEC,coldAt}. It has been also shown that is possible to
map the dynamics of such model on circuit QED~\cite{Circuit} and ion
traps~\cite{IonTraps} systems.  
\par
For what concerns metrology, the crucial feature of the LMG model 
is that its Hamiltonian depends on two parameters: one is the anisotropy
parameter, not accessible to the experimenter, while the other is the
external magnetic field, thus experimentally tunable, at least to some
extent, in order to drive the system towards criticality. 
\par
In this paper, we address quantum metrology in LMG systems.
We first consider the characterization of LMG systems, i.e. the
estimation of anisotropy, and show how criticality may be exploited to
improve precision. To this aim we evaluate exactly the quantum Fisher
information of small-size LMG chains made of $N=2,3$ and $4$ lattice
sites and also address the thermodynamical limit by a zero-th order
approximation of the system Hamiltonian. Our results show that the 
maxima of the quantum Fisher information are obtained on the critical
lines in the parameter space, i.e. where the ground state of the system
is degenerate. We also show that the ultimate
bounds to precision may be achieved in practice by tuning the external field and by
measuring the total magnetization of the system.  We also address the
use of LMG systems as quantum thermometers, i.e. we consider a LMG chain
in thermal equilibrium with its environment and analyze the estimation
of temperature by quantum-limited measurements on the sole LMG system.
We show that the precision is governed by the gap between the lowest
energy levels of the systems, such that the field-dependent level
crossing provides a metrological resource to extend the operating range of the
quantum thermometer. 
\par
The paper is structured as follows: in Section \ref{LMGModel} we briefly
review the relevant features of the LMG model in its most relevant forms, 
whereas in Section \ref{QET} we introduce the tools of quantum estimation
theory and establish notation. In Section \ref{QFIgamma} we analyze in details
estimation of anisotropy, whereas Section \ref{QFIbeta}
is devoted to LMG systems as quantum thermometers. A perturbation
analysis in order to discuss the robustness of the optimal estimators
against fluctuations of the external field is the subject of Section
\ref{hfl}. Finally, in
section~\ref{thermo} we address the thermodynamical limit by
means of a zero-th order approximation of the system Hamiltonian.  
Sec.~\ref{conclusion} closes
the paper with some concluding remarks.
\section{The LMG model}
\label{LMGModel}
In this section we review the main features of the Lipskin-Meshkov-Glick
model. As a matter of fact, the model has been widely studied in many
branches of science and it is known in several equivalent forms. We
present the most relevant ones, with emphasis on the symmetries of the
system.  
\par
The original formulation~\cite{LMG1,LMG2,LMG3}  describes a system of
$N$ fermions occupying two $N$-fold degenerated levels separated by an
energy gap $\epsilon$.  Let $s=-1,1$ be and index for the level an
$p=1,...N$ an index exploring the degeneracy of the levels, let us
consider a fermion algebra
$\left\{\alpha_{ps},\alpha_{p^{'}s^{'}}^{\dagger}\right\}=\delta_{p
p^{'}}\delta_{s s^{'}}  $ with $\alpha_{ps}$ (
$\alpha_{ps}^{\dagger}$) the annihilation (creation) operator of a
fermion in the $p-$th degenerated state of the $s$ level, then the
LMG Hamiltonian reads
\begin{align} \label{Hfer}
H = &\frac{\epsilon}{2}
\sum_{ps} s\, \alpha_{ps}^\dagger \alpha_{ps} 
+ \frac{\mu}{2}\sum_{pp's}\alpha_{ps}^\dagger 
\alpha_{p's}^\dagger \alpha_{p'-s}\alpha_{p-s} + \notag \\
&+\frac{\nu}{2}\sum_{pp's}\alpha_{ps}^\dagger 
\alpha_{p'-s}^\dagger \alpha_{p's}\alpha_{p-s}.
\end{align}
The first term takes into account the single-particle energies, the
second term introduces a scattering between couples of particles in the
same level and the third term is a level swapping for a couple of
particles with different $s$.
The model has the advantage of being simple enough to be solved exactly 
for small $N$ or numerically for large $N$. In fact, the symmetries of the 
system allows one to reduce the size of the largest matrix to be diagonalized. 
At the same time, the system is far
to be trivial, and allows one to test the goodness of many approximations 
techniques \cite{pan99,ros08}, as well to compare classical and quantum phase 
transitions \cite{cas06}.
\par
The Hamiltonian in Eq.(~\ref{Hfer}) may be rewritten in terms of angular 
momentum operators defined by 
\begin{equation} \begin{split}
S_z & =\frac{1}{2}\sum_{ps} s\, \alpha_{ps}^\dagger \alpha_{ps} \\
S_+ &= \sum_{p}^N \alpha_{p+1}^\dagger\alpha_{p-1} \qquad S_{-} = S_+^\dag 
\label{eq:spinfermioni}
\end{split} \end{equation}
and introducing new parameters 
\begin{equation} 
\nu  = -\frac{1}{N}(1+\gamma) \quad \mu  = \frac{1}{N}(1-\gamma)
\quad \epsilon  = -2 h \,,
\label{eq:paramsub}
\end{equation}
leading to \cite{uly92} (apart from an energy shift)
\begin{align}
H= &-\frac{1}{N}(1+\gamma)(S^2-S_z^2-\frac{N}{2}) \notag \\ 
         &-\frac{1}{2N}(1-\gamma)(S_+^2+S_-^2) - 2 h\, S_z \ .
\label{eq:LMGtot}
\end{align}
Finally, upon writing the $S$ operators as  collective spin operators 
$$S_\alpha \equiv \frac12 \sum_{k = 1}^N \sigma_\alpha^k\,,$$ we may rewrite the
LMG Hamiltonian as the Hamiltonian acting on the space of $N$
interacting spin $\frac12$ systems, also exposed to an external field, i.e.  
\begin{equation}
H = -\frac{1}{N}\sum_{j<k} (\sigma_x^j \sigma_x^k + 
\gamma \sigma_y^j \sigma_y^k)  - h \sum_k^N \sigma_z^k \  
\label{eq:HamLMG}
\end{equation}
where $\sigma_\alpha^k$ is the Pauli matrix associated to the 
direction $\alpha=x,y,z$ of the $k$-th spin. The sum is extended over
all the spins, thus describing a system where the interaction is not 
limited to first neighbours.
The first term in Eq.(~\ref{eq:HamLMG}) introduces a spin-spin
interaction whose strength is made anisotropic in the $xy$ plane by the
$\gamma$ parameter, which is the ratio between the coupling energies in
this directions ($\gamma =1$ means no anisotropy). Finally the strength
of the interaction with the external field is described by the parameter $h$.
\par
It is worth to point out some symmetries of the system. At first we
notice that the swap $h \rightarrow -h$ modifies the Hamiltonian as  the
(unitary) operations of describing  {\em spin flip}, i.e.  
$U = \bigotimes_{k=1}^N  \sigma_x^k$
\begin{equation}
H(\gamma,h) = U^{\dagger} H(\gamma,-h) U 
\label{eq:spinflip}
\end{equation}
so that there is no need to study the $h<0$ semi-plane, since the eigenvalues here are 
the same as in the $h>0$ case, and the eigenvectors are related by the transformation 
matrix $U$.  Similarly, the $\gamma$ parameter may be taken in the
range $[-1,1]$ since any map sending this range into $(-\infty,-1]$
$\bigcup$ $[1,\infty]$ modifies the Hamiltonian as a $\pi/2$ rotation around the 
$z$ axis, i.e. as the unitary $V = \bigotimes_{k=1}^N  \sigma_z^k  $
together with a rescaling of the field 
\begin{equation}
H(\frac{1}{\gamma},h) = V^{\dagger}\,H(\gamma,h  \gamma)\, V\,.
\label{eq:gammagreater1}
\end{equation}  
The parameter space is therefore
restricted to $(\gamma,h) \in [-1,1] \times [0,\infty)$.  
\par
The LMG model spectrum has been extensively studied in the
thermodynamic limit~\cite{vid1,vid2,vid3,rib07,rib08,bot1,bot2}. 
Following the method suggested 
in~\cite{vid1} the spectrum of $H$  in the large $N$ limit is computed using
first a Holstein-Primakoff bosonization
\begin{equation} \begin{split}
S_{+}& = N^{1/2}(1 - a^{\dagger}a/N)^{1/2} a \quad S_-= S_+^\dag \\ 
S_{z} & =N/2 - a^{\dagger}a, 
\label{HolsPrim}
\end{split} \end{equation}
and considering at most term in $(1/N)^{0}$ in the expansion of the square 
root. Subsequently in order to diagonalize $H$ a Bogoliubov transformation 
is performed  
\begin{equation} \begin{split}
\label{Bog}
a= \cosh \Theta\, b + \sinh \Theta\, b^\dag
\end{split} \end{equation}
where $\Theta\equiv\Theta(\gamma,h) $ is chosen such that the 
Hamiltonian reads (neglecting a constant energy shift)
\begin{equation} \begin{split}
H \stackrel{N\gg 1}{=}  \Delta(\gamma,h)\,b^{\dagger}b.
\label{eq:Hdiag}
\end{split} \end{equation}
The study of the ground state reveals two phases in the parameter space:
for $h\geq1$ the system shows an ordered phase with 
$$\Delta(\gamma,h) =2 \left[(h-1)(h-\gamma)\right]^{1/2}$$ while for $0 \leq h<1$ we
have a disordered (broken) phase with an energy spacing among levels given by 
$$\Delta(\gamma,h)=2 \left[(1-h^2)(1-\gamma)\right]^{1/2}\,.$$
\section{Quantum Estimation Theory}
\label{QET}
In this section we briefly review the basics of quantum estimation 
theory and the tools it provides to evaluate bounds to precision of
any estimation process involving quantum systems. 
Let us consider a situation in which the quantum state of a system is known 
unless for a parameter $\lambda$, e.g. a system with a known Hamiltonian in 
thermal equilibrium with a reservoir at unknown temperature $T$. This situation 
is described by a map $\lambda \rightarrow \rho_\lambda$ associating to each 
parameter value a quantum state. In this framework when one measures an observable 
$X$ the outcomes $x$ occur with a conditional probability distribution
$p_X(x|\lambda)$ given by 
\begin{equation}
p_X(x|\lambda) = \mathrm{Tr}[P_x \rho_\lambda],
\label{eq:probability}
\end{equation}
where $P_x$ is the projector onto the eigenspace relative to $x$.
In order to estimate the value of $\lambda$ from the data one needs an
{\em estimator}, i.e. a function $\hat{\lambda} \equiv \hat{\lambda} 
(x_1, x_2, ...) $ of the measurement outcomes to the parameter space.
Of course one requires some properties for this estimator, primarily 
to be unbiased
\begin{equation}
E[\hat{\lambda} - \lambda] = \prod_i \sum_{x_i}  \hat{\lambda}(x_1,...x_n) - \lambda = 0 
\qquad \forall \lambda,
\label{bias}
\end{equation}
where $E[.]$ denotes the mean with respect to the $n$ 
identically distributed random variables $x_i$ and $\lambda$ the true 
value of the parameter. Additionally one requires a small variance for the 
estimator 
\begin{equation}
\mathrm{Var}(\lambda,\hat{\lambda}) = E[\hat{\lambda}^2]-E[\lambda]^2\,,
\label{var}
\end{equation}
since this quantity measures the overall precision  of the inference process.
A lower bound for the variance of any estimator is given 
by the Cramer-Rao theorem 
\begin{equation}
\mathrm{Var}(\lambda,\hat{\lambda}) \geq \frac{1}{M F_\lambda},
\label{eq:CramerRao}
\end{equation}
where $M$ is the number of independent measurements and 
$F_\lambda$ is the Fisher Information (FI) given by 
\begin{equation}
F_\lambda = \sum _x  \frac{\left[\partial_\lambda
p_X(x|\lambda)\right]^2}{p_X(x|\lambda)}\,.
\label{eq:FI1}
\end{equation}
An estimator achieving the Cramer-Rao bound is said to be efficient.
Although an efficient estimator may not esist for a given data set, in
the limit of large samples, i.e. for $M\gg1$, an asymptotically
efficient estimator always exist, e.g. maximum likelihood estimator. 
In summary, once a map $\lambda \rightarrow \rho_\lambda$ is given it is possible to infer 
the value of a parameter of a system by measuring an observable $X$ and performing statistical 
analysis on the measurements results. Upon choosing a suitable
estimator we may achieve the optimal inference, i.e. saturate at least asymptotically 
the Cramer-Rao bound.
\par
It is clear that different observables lead to a different probability distribution, giving 
rise to different FIs and hence to different precisions for the estimation of $\lambda$~\cite{LQE}. 
The ultimate bound to precision is obtained upon maximizing the FI over the set of observables. This maximum is the so-called quantum Fisher information (QFI). In order 
to obtain an expression for the QFI one introduces the symmetric logarithmic derivative
(SLD), which is the operator $L_\lambda$ solving 
\begin{equation}
\frac{L_\lambda \rho_\lambda +  \rho_\lambda L_\lambda}{2} =
\frac{\partial \rho_\lambda}{\partial \lambda}\,.\label{eq:SLD}
\end{equation}
SLD allow us to rewrite the derivative of $\rho_\lambda$ so 
that Eq.(~\ref{eq:FI1}) becomes 
\begin{equation}
F_\lambda = \sum_x \frac{\mathrm{Re}(\mathrm{Tr}[\rho_\lambda 
P_x L_\lambda])^2}{\mathrm{Tr}[\rho_\lambda P_x L_\lambda]}\,,
\label{eq:FI2}
\end{equation}
which is upper bounded by 
\begin{equation}
F_\lambda \leq \mathrm{Tr}[\rho_\lambda L_\lambda ^2] \equiv G_\lambda.
\label{eq:optimiz}
\end{equation}
where $G_\lambda$ is the quantum Fisher information.
In order to obtain an explicit form for the QFI one has to solve
Eq.(\ref{eq:SLD}), arriving at 
\begin{equation}
L_\lambda = 2 \int_0^\infty\!\!\! dt\, e^{-\rho_\lambda t}\, \partial_\lambda
\rho_\lambda\, e^{- \rho_\lambda t}\,.
\label{eq:SLD12}
\end{equation}
Then, upon writing 
$\rho_\lambda = \sum_n w_n(\lambda) |\psi_n(\lambda)\rangle \langle \psi_n(\lambda) | $ in 
its eigenbasis, we have
\begin{equation}
L_\lambda = 2 \sum_{n m} \frac{\langle 
\psi_n|\partial_\lambda \rho_\lambda |\psi_m \rangle}{w_n + w_m} |\psi_n 
\rangle \langle \psi_m |\,,
\label{eq:SLD13}
\end{equation}
and finally
\begin{equation}
G _\lambda = 2 \sum_{n m} \frac{|\langle 
\psi_n|\partial_\lambda \rho_\lambda |\psi_m \rangle|^2}{w_n + w_m} ,
\label{eq:QFI2}
\end{equation}
with the sum carried over those indexes for which $w_n + w_m \neq 0$.
Upon rewriting $\partial_\lambda \rho_\lambda $ in terms of 
the eigenvectors and the eigenvalues of $\rho_\lambda $, we have
\begin{equation}
\partial_\lambda \rho_\lambda = \sum_n \partial_\lambda w_n 
|\psi_n\rangle \langle \psi_n | + w_n |\partial_\lambda 
\psi_n\rangle \langle \psi_n | + w_n |\psi_n\rangle \langle \partial_\lambda \psi_n | ,
\label{eq:drho}
\end{equation}
and the QFI assumes the following form
\begin{equation}
G _\lambda = \sum_n \frac{(\partial_\lambda w_n )^2 }{w_n} 
+ 2 \sum_{n \neq m} \sigma_{nm} | \langle \psi_n |\partial_\lambda
\psi_m \rangle |^2\,,
\label{QFIfinal}
\end{equation}
with 
\begin{equation}
\sigma_{nm} = \frac{(w_n - w_m)^2}{w_n + w_m}.
\label{eq:sigma}
\end{equation}
The first contribution in the Eq. (\ref{QFIfinal}) depend solely on 
the eigenvalues of  $\rho_\lambda$, i.e. on the fact that 
$\rho_\lambda$ is a mixture, whereas the second term depends on 
the eigenvectors, i.e. it contains the truly quantum contribution to QFI. 
The two terms are usually referred to as the {\em classical} and the 
{\em quantum} contribution to the QFI, respectively. For pure states 
the quantum term is the only one contributing to the QFI.
\section {Estimation of anisotropy}
\label{QFIgamma}
The interaction described by the LMG model depends on two relevant 
parameters: the anisotropy $\gamma$ and the external field $h$.  To 
these it adds the temperature, or equivalently its inverse $\beta$, if 
we allows the system to interact with the environment by exchanging 
energy. Among these parameters, the external field may be tuned by the 
experimenter and represents a tool that allows ones to exploit the 
systems criticality as a resource to reliably estimate the remaining 
less controllable parameters.
\par
The anisotropy is a typical quantum parameter, that is its variations
modify both the eigenvalues and the eigenvectors of the system. Anisotropy 
is not tunable by the experimenter, since it is part of the intrinsic coupling 
among spins and represents a specific characteristic of the system. Anisotropy
however {\em does not} correspond to a proper observable. Its
characterization may be addressed within the framework of QET and the
ultimate bound to the precision of its estimation is set by the
corresponding QFI. 
\par
We consider here LMG chains in thermal equilibrium with their environment. 
The map that we mentioned in the previous Section, from parameters space to 
quantum states is thus given by the canonical Gibbs density matrix
\begin{equation}\begin{split}
\rho(\gamma,h, \beta ) &= 
\frac{e^{- \beta H(\gamma,h)}}{Z(\gamma,h,\beta)} \\ 
&= \sum_n \frac{e^{-\beta E_n (\gamma,h)} }
{Z(\gamma,h,\beta)}\,|n\rangle \langle n|\,,
\label{eq:BoltzStates}
\end{split}\end{equation}
where $Z(\gamma,h,\beta)= \mathrm{Tr} [e^{-\beta H} ]$ is the 
partition function, $E_n(\gamma,h)$ the $n$-th eigenvalue of 
the Hamiltonian, and $| n\rangle$ a basis where $H$ is diagonal, such
that $\rho(\gamma,h, \beta )$ has eigenvalues equal to the Boltzmann weights   
\begin{equation}
B_n \equiv B_n(\gamma,h,\beta) =
\frac{e^{-\beta E_n (\gamma,h)} }
{Z(\gamma,h,\beta)}
\label{eq:pesiboltz}
\end{equation} 
\par
In order to evaluate the QFI for $\gamma$, an in turn the bounds to
precision in its estimation, we have to find the eigenvalues and 
eigenvector as a function of $\gamma$ and $h$ and insert them in 
Eq. (\ref{QFIfinal}). To gain some insight into the role
of the chain size while maintaining the approach analytic,
we have analyzed in details the cases $N=2,3,4$. We will address 
the complementary limit $N \to \infty$ in Sec.\ref{thermo}.
\par
Before proceeding with the results we take a preliminary observation: by
studying parameter estimation through information geometric tools such
as the QFI and the FI one learns that the parameter of interest is easy
to estimate in those points where the parametrized quantum state is
easily distinguishable (in a statistical sense) from the neighboring
ones, corresponding to slightly different values of the parameter. In
our case, upon looking at the very form (\ref{eq:BoltzStates}) of the
quantum state, one sees that for small values of $\beta$ $\rho$ is
almost independent by $\gamma$, going toward a uniform mixture of all
the eigenstates. In this regime, one thus expects the estimation of
$\gamma$ to be inherently  inefficient.  On the other hand, high
precision is expected in the large $\beta$ limit, since the mixture is
peaked at the ground state, which is intuitively more sensitive on
$\gamma$ fluctuations. 
\par
Using Eq.(\ref{QFIfinal}) and the results of diagonalization 
(see Appendix \ref{a:DD}),
one arrives at the QFI $G_\gamma\equiv G_\gamma(\gamma,h,\beta)$. 
For $N=2$ the explicit expression is given by 
\begin{align}
G_\gamma = \frac{1}{r^2}\left[\beta^2\, \frac{\kappa_1}{2\kappa_2} + 
\frac{16 h^2}{r^2} \frac{(1-e^{\beta r})^2}{(1+e^{\beta
r})\sqrt{\kappa_2}}
\right]\,, \label{gopt2}
\end{align}
where 
\begin{align*}
\kappa_1=&\, e^{-\frac12 \beta (v - r)} 
\Big[\frac12 (u - r)^2 + 4 (8 h^2 + u^2) e^{\frac12 \beta (v + r)}
\notag \\
&+ \frac12 (u - r)^2  e^{\beta  (v + r) } + 
\frac12 (u + r)^2  e^{\beta r}+ 
\frac12 (u + r)^2  e^{v \beta} \Big] \notag \\
\kappa_2=&\left[1 + e^{\beta r} + e^{\frac12 \beta (v + r)} + e^{-\frac12 \beta
(v - r)}
\right]^2
\end{align*}
with $u=\gamma-1$, $v=\gamma+1$ and $r=\sqrt{u^2+16 h^2}$.
For $N=3$ and $N=4$ the expressions are quite cumbersome and we are
not reporting them.
\par
Optimal estimation of the anisotropy at fixed temperature 
may be achieved by maximizing the QFI over the external field $h$.
Results of this maximization show that the optimal values 
of the field correspond to the {\em critical lines} of the model, 
i.e. the lines in the parameter space correspoding to a 
degenerated ground state (GS), i.e. 
\begin{align}
N&=2 \rightarrow h_c=\frac{ \sqrt{\gamma} }{2}\\ 
N&=3 \rightarrow h_c=\frac{2 \sqrt{\gamma}}{3} \\ 
N&=4 \rightarrow h_c=\frac{ \sqrt{\gamma}}{4} \hbox{  and  } 
h_c=\frac{ 3\sqrt{\gamma}}{4}\,.
\label{clines}
\end{align}
For $N=2$ the maximized QFI $G_\gamma (\gamma,\frac12
\sqrt{\gamma},\beta)$ is given by
\begin{align}
G_\gamma^{\rm opt} = \frac{8\gamma + \kappa^2 + \gamma (\gamma\kappa^2-8
)\, {\rm sech}^2 \frac 12 \kappa}{4 (1+\gamma)^4}
\label{ggopt}
\end{align}
where $\kappa = \beta (1+\gamma)$.
In the low temperature regime, i.e. $\beta \gg 1$ we may write
\begin{align} \label{ggasy}
G_\gamma^{\rm opt} \simeq \beta^2  \frac{(u + r)^2}{8 r^2} \left\{
     \begin{array}{lr}
            e^{\frac12 \beta (v -  r)} &  h \geq \sqrt{\gamma}/2 \\
	           e^{-\frac12 \beta (v-r)}  &  h < \sqrt{\gamma}/2
		        \end{array}
			   \right.\,.
			   \end{align}
Notice that the exponent is the energy gap between the two lowest energy
eigenvalues, which vanishes on the degeneracy line.
For $N=4$ the absolute maximum corresponds to 
$h_c=\frac{ 3\sqrt{\gamma}}{4}$. For $N=3$ also the condition 
$h=0$ individuates a degenerated GS, but this is not 
corresponding to a maxima of the QFI for reasons that will be clear
in the following. 
\par
\begin{figure}[ht!]
\includegraphics[width=0.48\columnwidth]{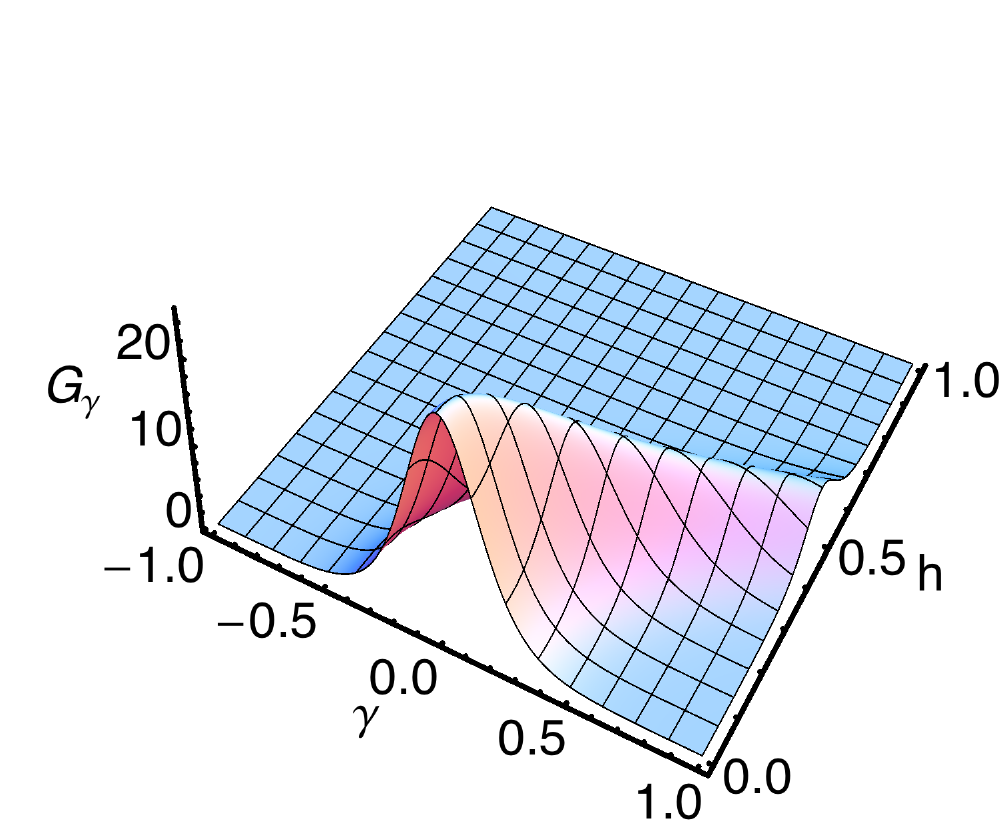}
\includegraphics[width=0.49\columnwidth]{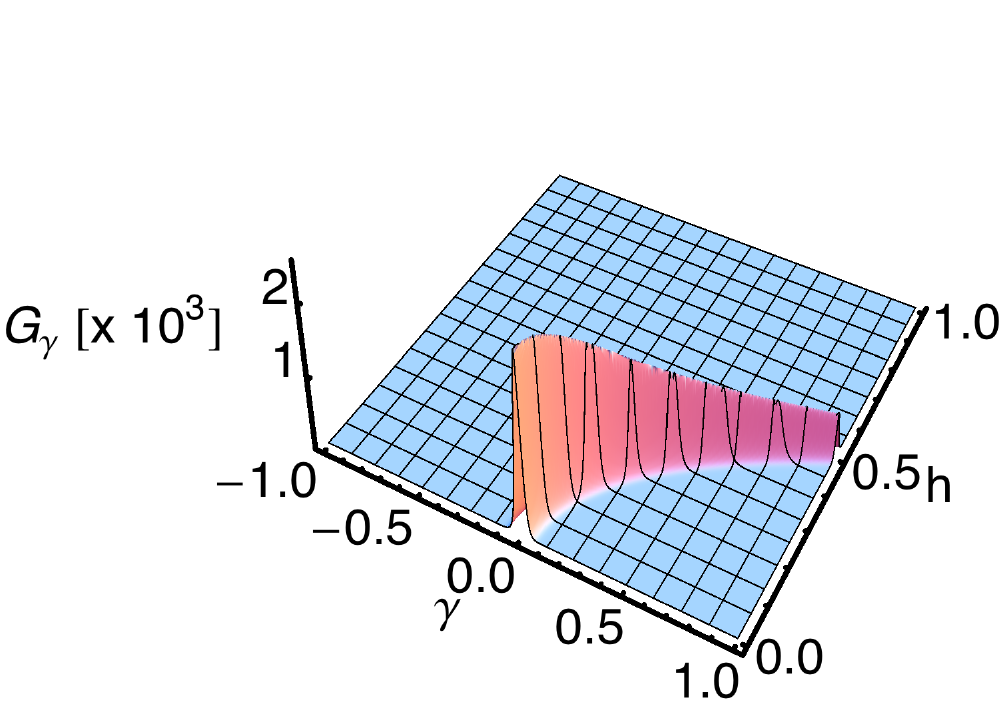}
\includegraphics[width=0.48\columnwidth]{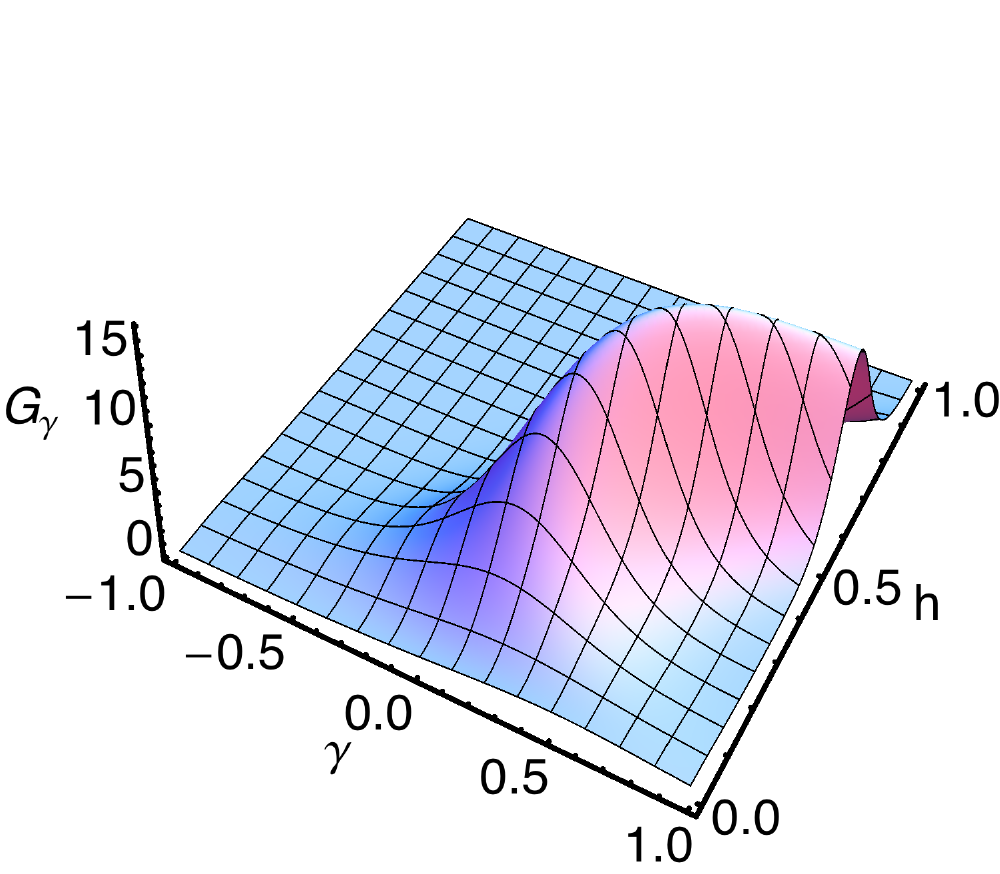}
\includegraphics[width=0.49\columnwidth]{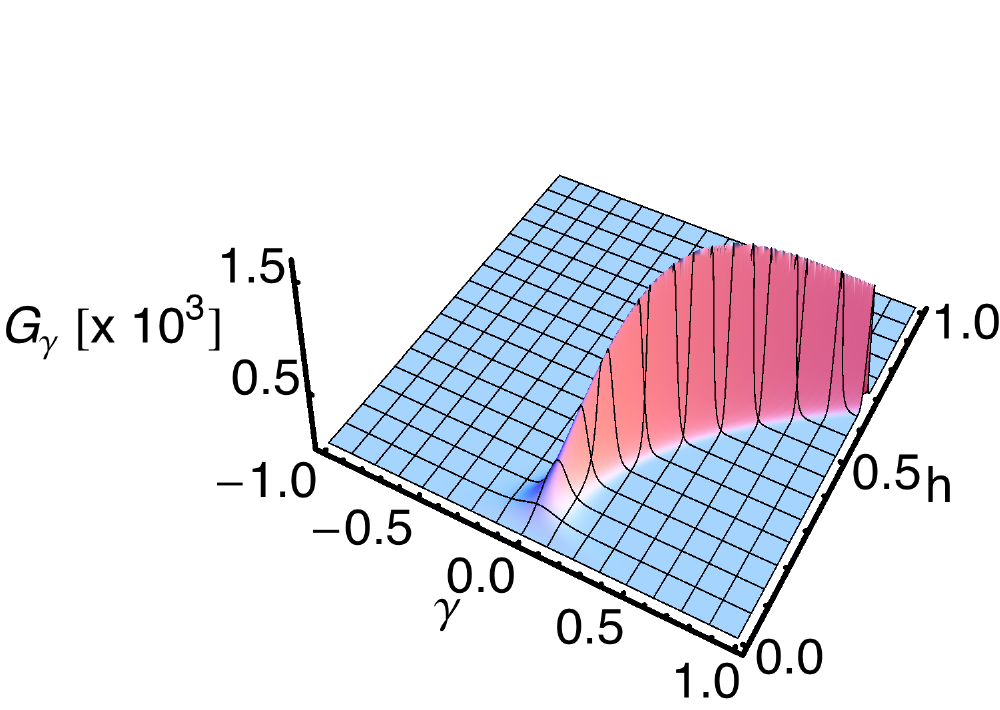}
\includegraphics[width=0.48\columnwidth]{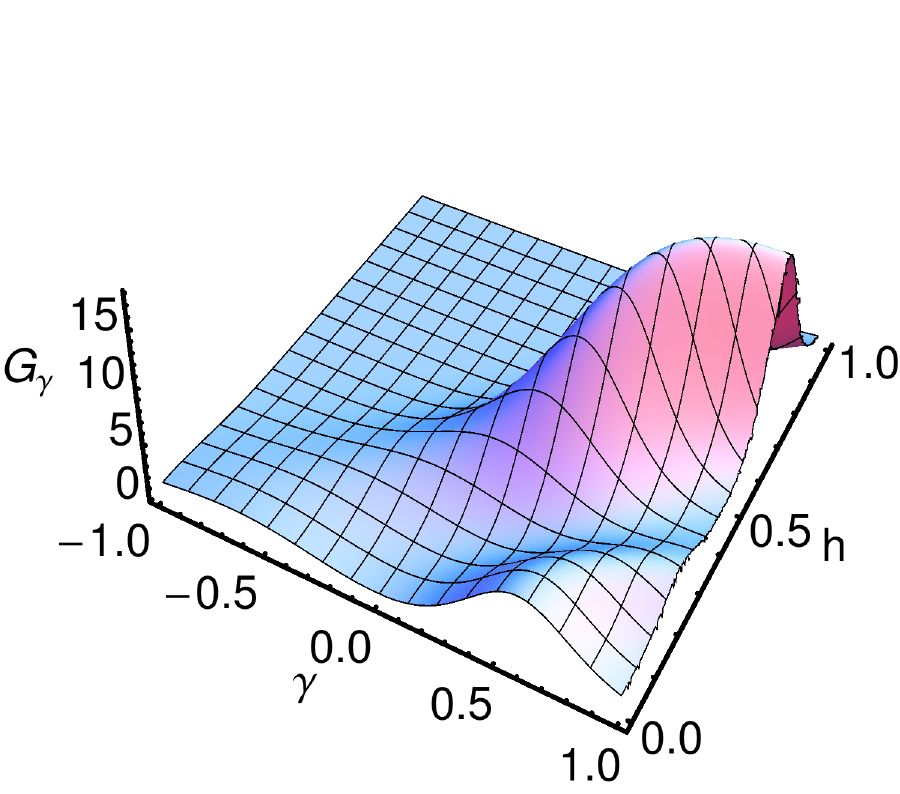}
\includegraphics[width=0.49\columnwidth]{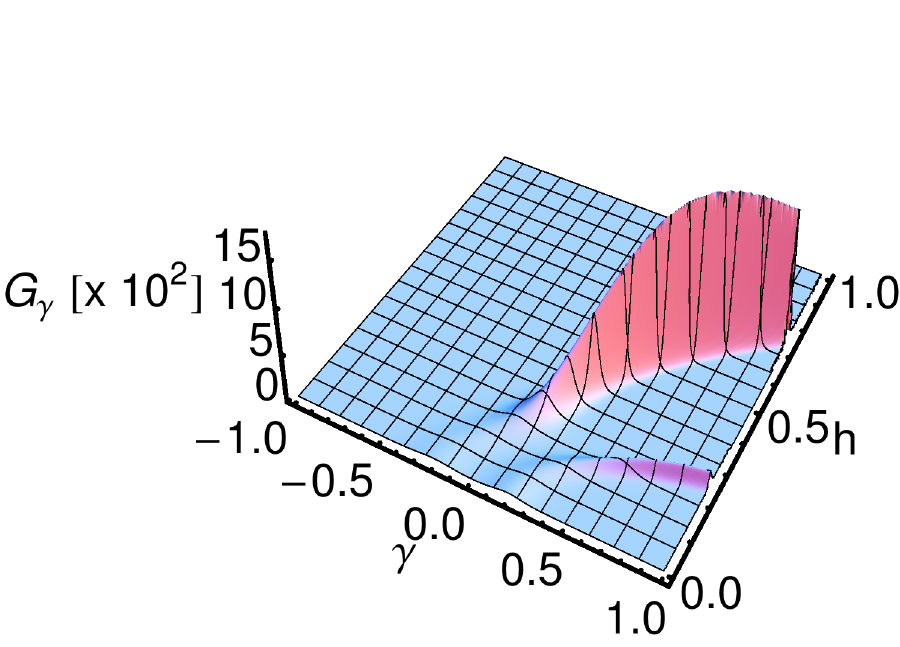}
\caption {Estimation of anisotropy in the LMG model. The plots show 
the QFI $G_\gamma$ for the anisotropy as a function of the anisotropy
parameter $\gamma$ itself and of the magnetic field $h$ for two values
of $\beta$. The panels of the left refer to $\beta=10$ while those on
the right to  $\beta=100$. The rows, from top to bottom, contain the
results for $N=2,3,4$ lattice sites respectively. Comparing the two
columns it is clear that $G_\gamma$ reaches its maximum along the
critical lines of the system as $\beta^2$, with such divergence
modulated also by a non trivial function of $\gamma$. Note the peculiar
absence of divergence in the N=3 case for h=0.  \label{fig:firstpanel}}
\end{figure}

The role of criticality is illustrated  in details in 
Fig.~\ref{fig:firstpanel}, where we show $G_\gamma$ as a 
function of $\gamma$ and
$h$ for different values of $\beta$. As it is apparent from the plots, 
when the temperature decreases $G_\gamma$ diverges as $\beta^2$ on the 
critical lines, whereas in any other point of the parameter 
space it assumes a finite value. In other words, for any value $\gamma
\geq 0$ it is possible to tune the external field to an optimal value
which drives the system into the degeneracy lines, i.e. into critical
points. In this way, one maximizes the QFI and, in turn, optimize
the estimation of $\gamma$. This results confirm that criticality is in
general a resource for estimation procedures. The degeneracy line at
$h=0$ line for $N=3$ is an exception, since no gain in precision is
achieved despite a crossing between the two lowest energy states is present. We will
address this issue and clarify the point in the following subsection.
\subsection{A two-level approximation to assess estimation of anisotropy 
in the low temperature regime}
\label{Qaprox}
An intuitive understanding of our findings may be achieved by
means of an approximation for the Gibbs states, where we 
consider only the two lowest levels of the system
\begin{equation}
\rho(\gamma,h,\beta) \propto 
e^{-\beta E_0} |0\rangle\langle 0| +
e^{-\beta E_1} |1\rangle\langle 1|
\label{apx}
\end{equation}
where $E_{0,1}$ are the smallest eigenvalues. 
In fact, for the values of $N$ we have considered, the energy 
spectra of the Hamiltonians show a common structure: the two lowest 
eigenvalues, i.e. the GS and the first excited level, cross each other 
but they remain smaller than the other levels for the whole range of  
$\gamma$ and $h$ values. As a consequence, for large $\beta$ (i.e. in the low 
temperature regime) the
Boltzmann weigths corresponding to the smallest eigenvalues are the only
appreciable in the sum in Eq.(~\ref{eq:BoltzStates}) and the density
matrix is well be approximated by the expression in Eq. (\ref{apx}).
The approximation is more and more justified as far as $\beta$
increases. 
We now proceed by noticing that for the family of 
states (\ref{apx}), the quantum contribution to $G(\gamma)$ 
does not contain any divergent term in $\gamma$, $h$ or $\beta$.
This may be easily seen from Eq. (\ref{QFIfinal}) and 
from the fact that the eigenvectors are smooth functions 
of the parameters. Actually, this is the case also for other
first-neighbour models \cite{ZP,Inv1}, so that the
approximation here described may apply to other models.
We thus introduce a general notation in order to analyze the 
classical contribution.  
\par
Consider a qubit with eigenenergies  
$f(a,b)$ and $g(a,b) = f(a,b) + x(a,b)$ , depending on the 
parameters $a$  and $b$ ($b$ may also be a {\em set} of parameters).
With the usual map to the thermal state, the QFI for the parameter 
$a$ rewrites
\begin{align}
G_a(a,b,\beta)=&\beta^2\, 
\frac{e^{\beta x(a,b)}} {\left[1+e^{\beta x(a,b)}\right]^2}
\left[\partial_a x(a,b)\right]^2
\label{eq31}
\end{align}
It is easy to see that $G_a(a,b,\beta)$ diverges only in those points
$a_0$ and $b_0$ such that $f(a_0,b_0) = g(a_0,b_0)$ and
$\partial_a f(a_0,b_0) \ne \partial_a g(a_0,b_0)$.  When this happens, QFI 
is proportional to $\beta^2$. The two conditions are indeed satisfied
on the degeneracy lines mentioned above, except for the case $N=3$ and
$h=0$, where the partial derivatives of the eigenvalues are equal, thus
preventing the divergence of the QFI.
\subsection{Achieving the ultimate bound to 
precision using feasible measurements}
\label{QFImag}
In the previous Sections we have evaluated the ultimate bound to
precision for the estimation of anisotropy, and have shown that the
level crossing driven by the magnetic field is a resource for the
estimation. In order to exploit this {\em quantum critical} enhancement
one has in principle to implement the measurement of the symmetric
logarithmic derivative which, in turn, should be an accessible
observable for the LMG system under investigation.  Since it is unlikely
to have such a control on a quantum system that any observable is
measurable, one is generally led to assess estimation procedure based on
realistic observables, i.e. to evaluate their Fisher Information and to
compare this function with the QFI.  
\par
In this section we consider a realistic observable, the total
magnetization of the LMG system, and compute the corresponding FI for
the estimation of anisotropy.  As we will see, this quantity approaches
the QFI in the critical region, thus showing that quantum critical
enhancement of precision is indeed achievable in an an experimentally
accessible scenario. 
\par
The total magnetization is diagonalized in the basis $\otimes_{k=1}^{N}
\ket{m_z}_k$, where $m_z \in {1,-1}$ and $\ket{x}_k$ denotes the
eigenvectors of the $z$ spin component of the k-th spin. If $N_z$ is the
number of spins up for a given basis element, the corresponding
eigenvalue is simply $\sum_{k=1}^N i = 2 N_z - N$, and the probability
of such measurement outcome, with the notation of Eq.(~\ref{eq:FI1}), is
given by
\begin{equation}
p(2 N_z - N, \lambda) = \frac{\mathrm{Tr} 
\left[ P_{N_z}\, \exp{-\beta H} \right]}{Z},
\label{eq:probmag}
\end{equation}
where $P_{N_z}$ denotes the projector onto the subspace spanned by the basis elements 
with $N_z$ spins up. Finally, to compute the corresponding Fisher Information $F_\gamma$ 
we substitute these probabilities in Eq.~(\ref{eq:FI1}). 
\par
We are not going to report the explicit formula for the $F_\gamma$, which is quite 
unhandy. Rather, we introduce and discuss an approximation which allows us to
reproduce its main features. We anticipate that $F_\gamma$ shares with the QFI the nice 
behaviour in the critical region, i.e. it diverges as $\beta^2$ on the degeneracy lines, 
except for the case $h=0$ line for $N=3$.
\par
Let us  consider a two-dimensional system prepared in the mixed state 
 $\rho(\lambda) = p \ketbra{0}{0} + (1-p) \ketbra{1}{1}$ where both the eigenvalue 
$p$ and the eigenvectors are functions of a parameter $\lambda$ to be estimated.
If a measurement of an observable $A = x_1 \ketbra{x_1}{x_1} + x_2 \ketbra{x_2}{x_2}$ 
is performed, the outcomes are distributed according to 
\begin{equation*}
P(x_i) = \mathrm{Tr}[\rho \ketbra{x_i}{x_i}] = p |\braket{0}{x_i}|^2 + (1-p) |\braket{1}{x_i}|^2\,,
\end{equation*}
where taking into account the normalization of the basis involved, we have the 
following relations
\begin{align}
q = & |\braket{0}{x_1}|^2 =  |\braket{1}{x_2}|^2 \\
1-q = & |\braket{0}{x_2}|^2 = |\braket{1}{x_1}|^2 
\label{eq:simplifications}
\end{align}
We will also denote with $\delta q = q - (1-q)$ and $\delta p = p - (1-p)$.
With this notation the FI for $A$ is rewritten in a compact form as
\begin{equation}
\mathrm{F}(\lambda) = \frac{ (\partial_\lambda p \ \delta q + \partial_\lambda q \ 
\delta p)^2}{(p \ \delta q -q) (p \ \delta q + 1 -q)}
\end{equation}
Specializing this to the case of our interest we have $1 - p = \exp(-\beta \epsilon)/Z $ where $\epsilon = \epsilon(\gamma,h)$ denotes the energy of the first excited level. Without lost of generality we can assume the energy of the GS is to be null, we thus arrive at 
\begin{align}
\partial_\gamma p &= \frac{\beta\, e^{\beta \epsilon}}{\left[1+e^{\beta\epsilon}\right]^2}  \partial_\gamma \epsilon\,.  
\label{derp}
\end{align}
Eq.  (\ref{derp}) implies that the FI $F_\gamma$ of any observable of the form 
$A = x_1 \ketbra{x_1}{x_1} + x_2 \ketbra{x_2}{x_2}$  diverges as 
$\beta^2$ in the large $\beta$ limit, provided that $\delta q \neq 0$ (this means 
that the two eigenstates must be distinguishable by that measurement), 
$\partial_\gamma \epsilon \neq 0$ (similarly to what we found for the QFI) and 
that $\epsilon = 0$, i.e. that we are at a critical point. Notice that the above model, 
basically the same we used to explain the results obtained
for the QFI, is valid to discuss the estimation performances of the total magnetization, but
cannot be used to approximate the FI of {\em any} observable $A$ of the LMG 
model in the limit of low temperature. In fact, even though the state of the sistem may be always 
approximated by a qubit, there is no reason for a general observable to be approximated 
by an operator acting in the qubit space only. 
\section{LMG critical systems as quantum thermometers}
\label{QFIbeta}
In this section we explore the performances of LMG critical systems 
as quantum thermometers, i.e. we consider a LMG systems in thermal 
equilibrium with its environment and analyze
the estimation of temperature by quantum-limited measurements on
the sole LMG system. In other words, we address the estimation of the 
temperature, viewed as an unknown parameter of the Gibbs distribution,
on the family of states defined in Eq.~(\ref{eq:BoltzStates}) \cite{bru11,bru12}.
\par
\begin{figure}[h!]
   \includegraphics[width=0.47\columnwidth]{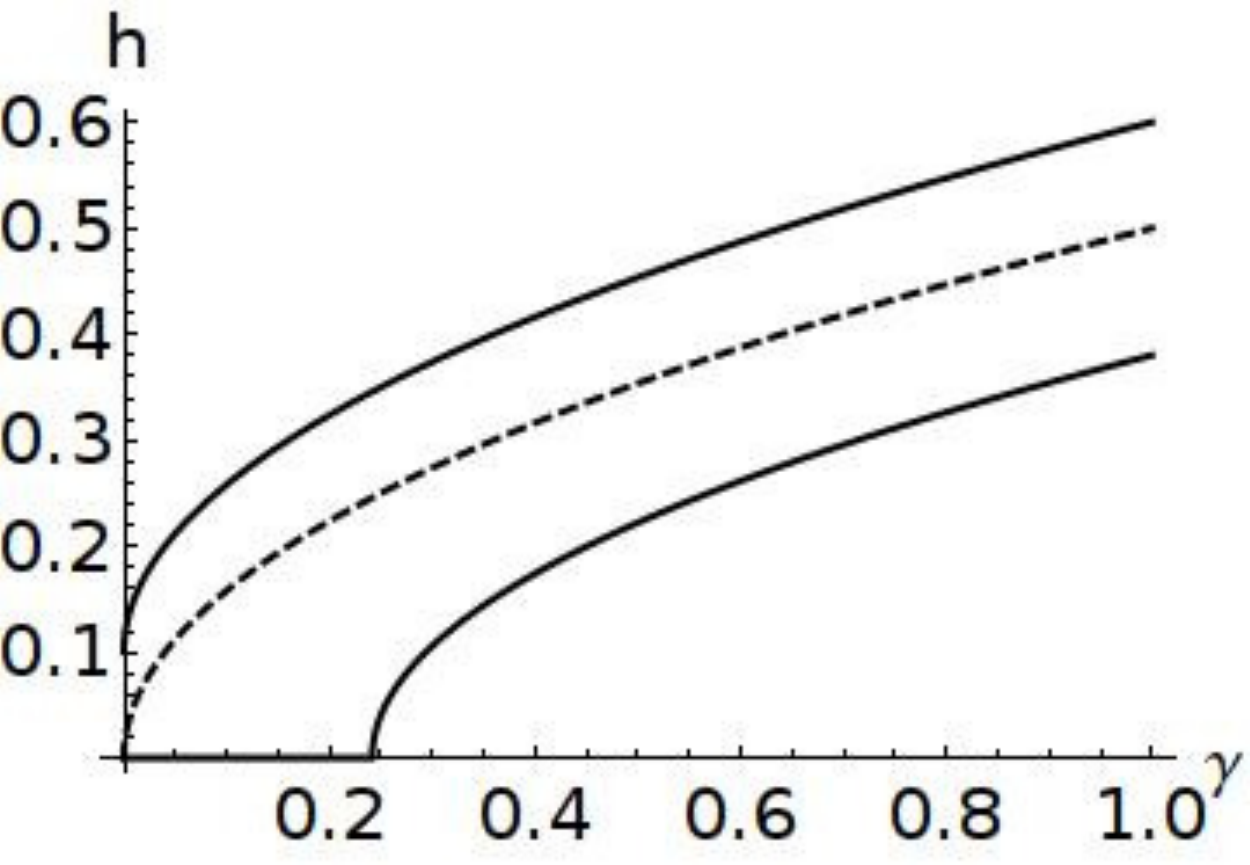}
  \includegraphics[width=0.47\columnwidth]{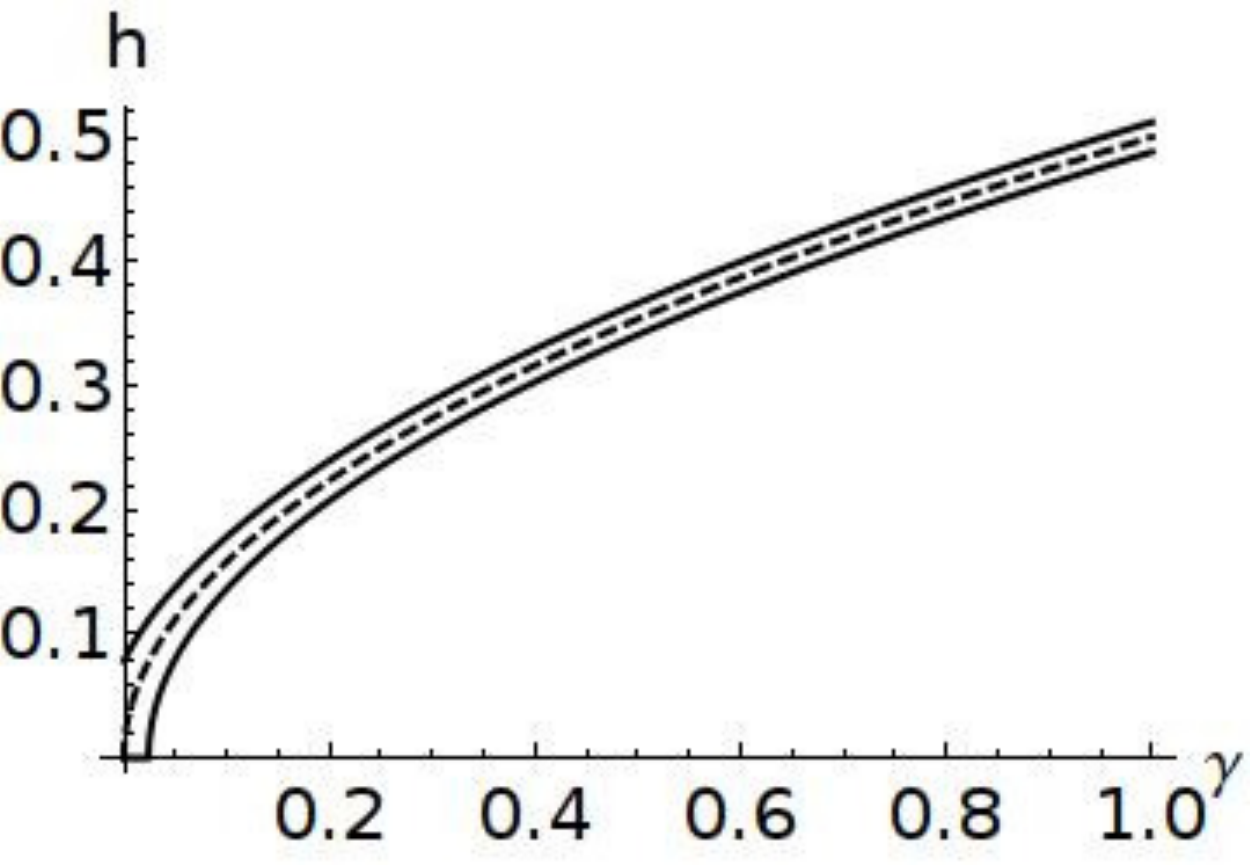}
  \includegraphics[width=0.47\columnwidth]{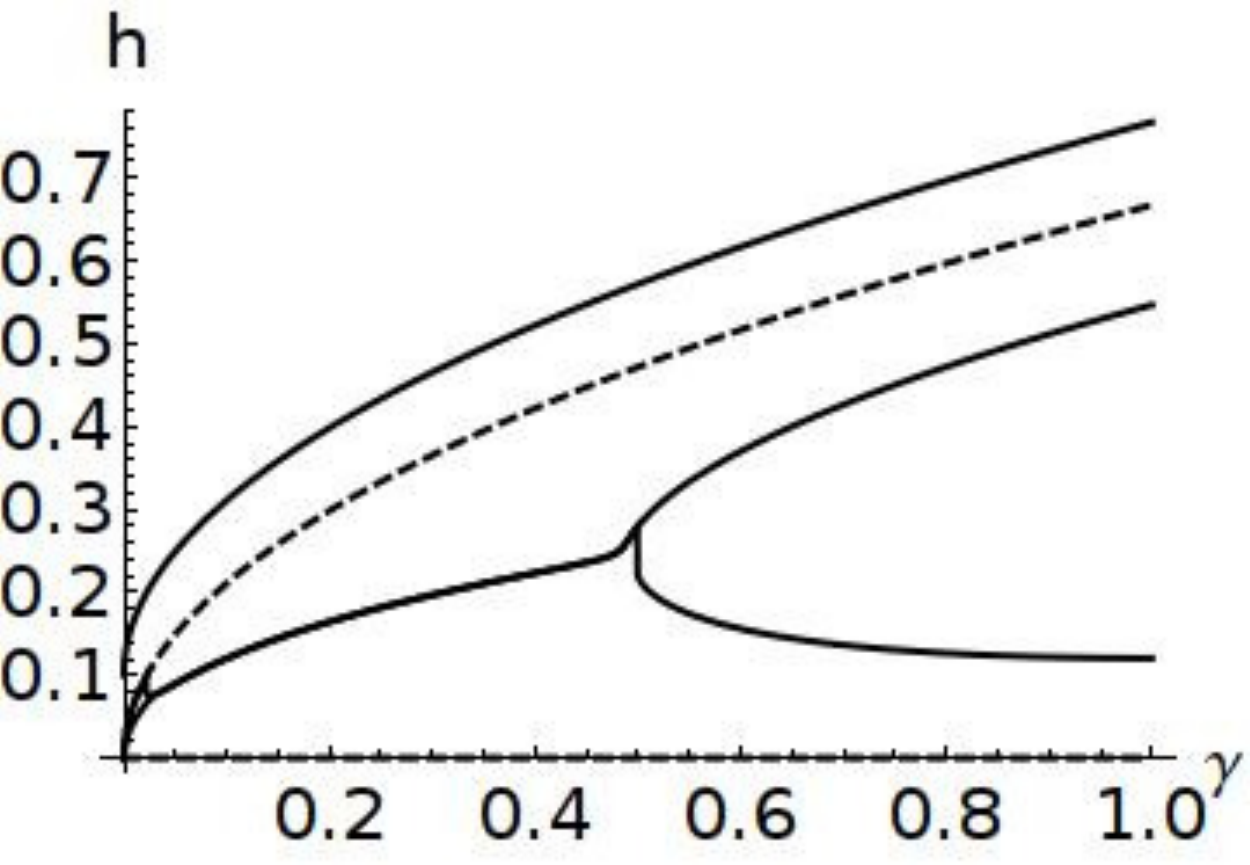}
  \includegraphics[width=0.47\columnwidth]{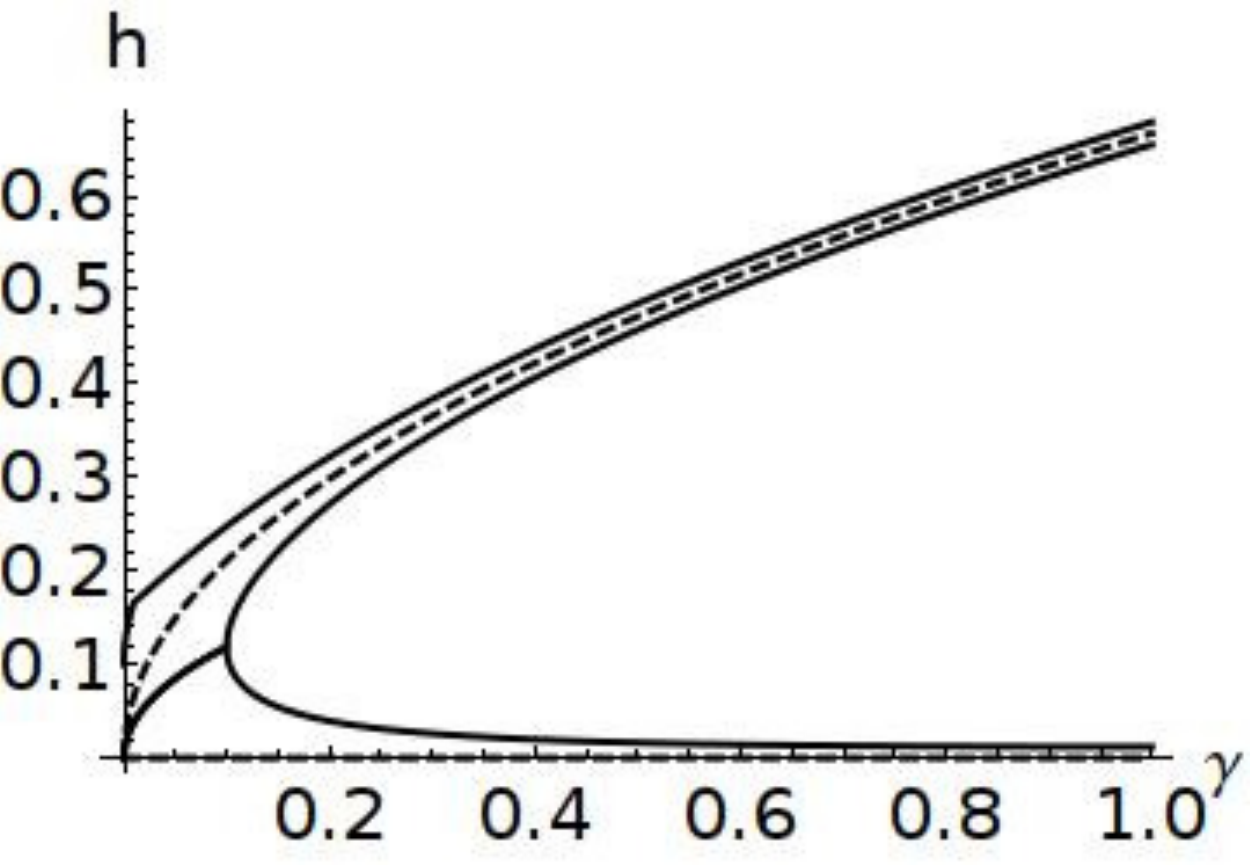}
  \includegraphics[width=0.47\columnwidth]{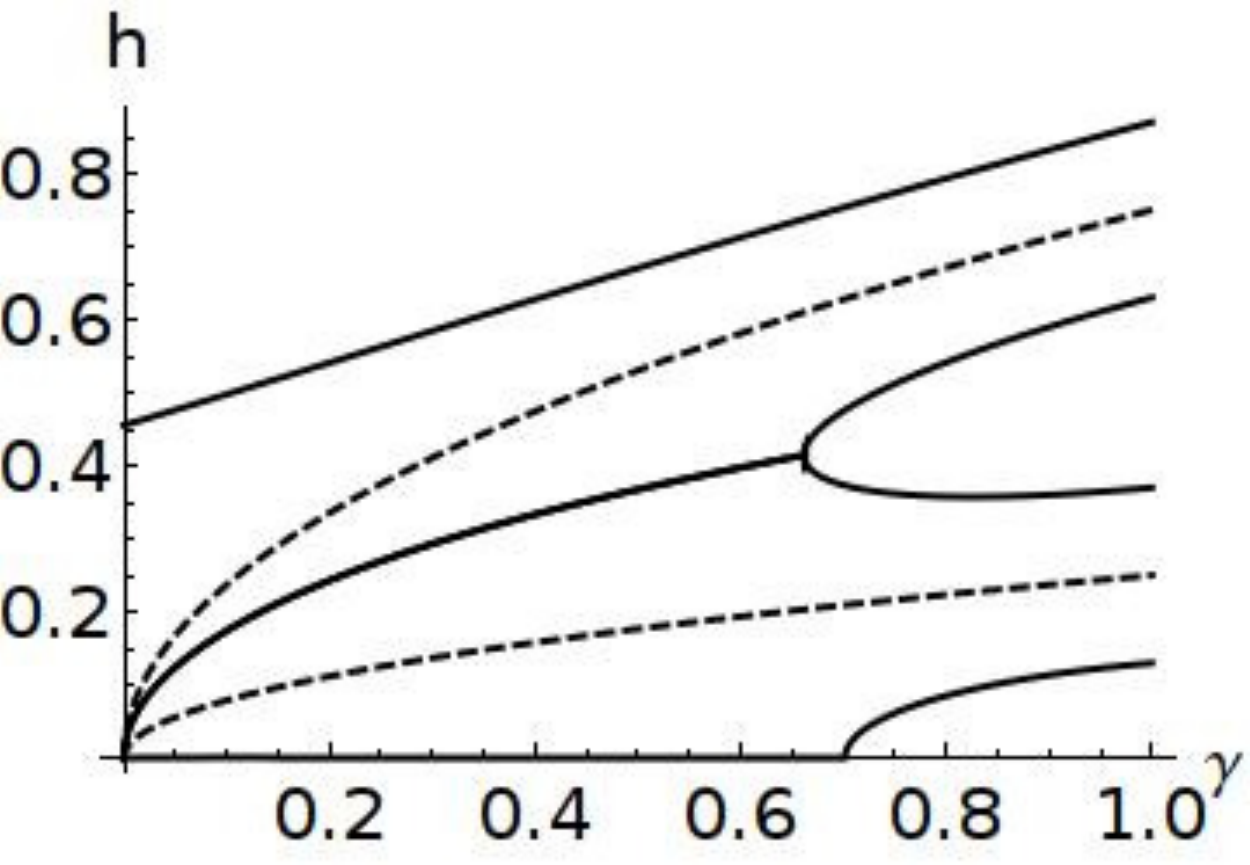}
  \includegraphics[width=0.47\columnwidth]{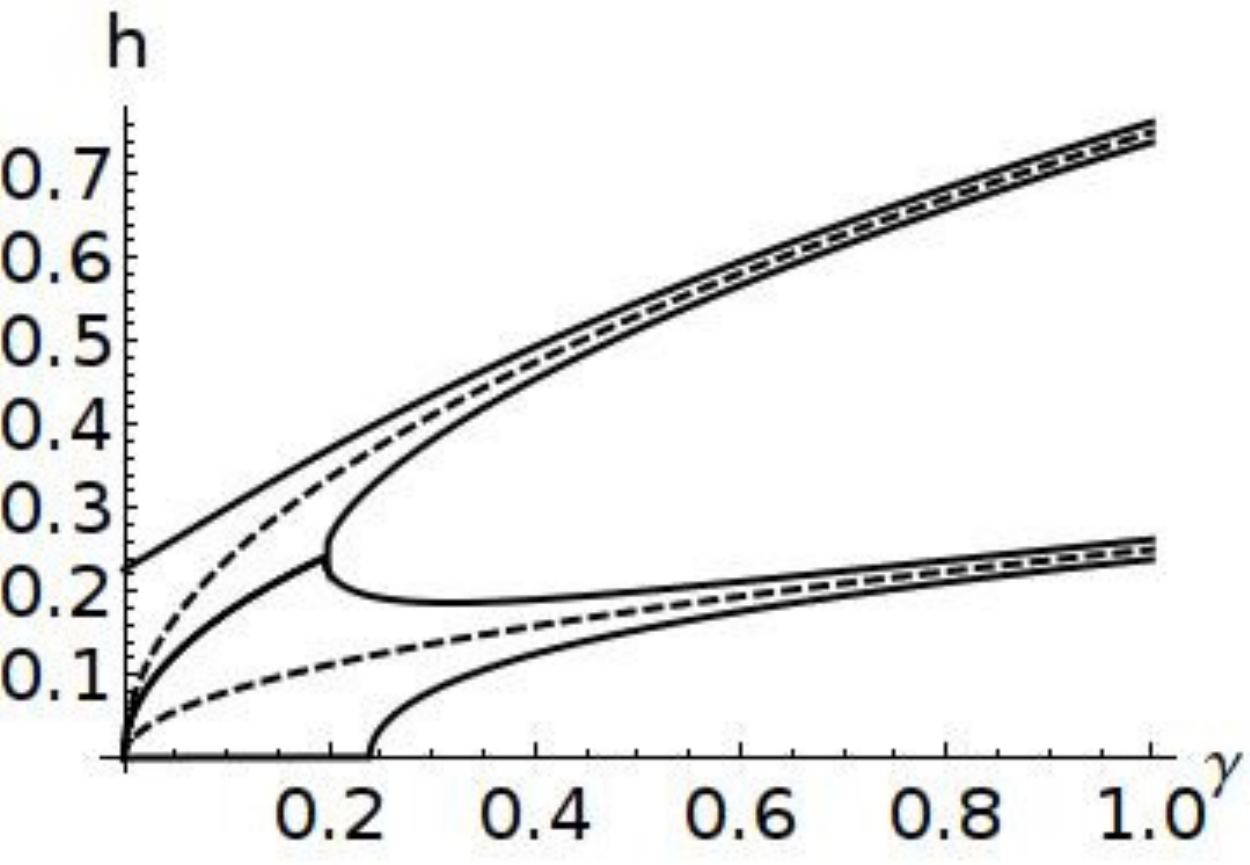}
\caption {Quantum thermometry using LMG systems. The plots show the
optimal field $h^*$, maximizing the QFI $G_\beta$, as a function of the
anisotropy of the system for different values of $\beta$ and for
different lengths of the LMG chain. Each row show the optimal field
versus $\gamma$ at fixed value of $N=2,3,4$ respectively. The two
columns correspond to $\beta=10$ and $\beta=100$ respectively. The
optimal values of the field are the solid lines whereas the dashed lines
are the critical lines $h_c$ of Eq. (\ref{clines}).} \label{fig:234termoptf}
\end{figure}

Upon inspecting Eq.~(\ref{eq:BoltzStates}) one easily sees that temperature 
influences the eigenvalues of the density matrix, but  not its eigenvectors, 
and thus only the classical contribution to the QFI $G(\beta)$ survives,
i.e. the sum depending on the Boltzmann weights in the general expression 
for QFI of Eqs.(~\ref{QFIfinal}). We thus have
\begin{equation}
G_\beta(\gamma,h,\beta)=\sum_{n=1}^d \frac{(\partial_{\beta}B_n)^2}{B_n},
\label{eq:QFItemp}
\end{equation}
where $B_n$ denotes the n-th Boltzmann weight. It is worth underlining that $G_\beta(\gamma,h,\beta)$ is 
equal to the energy fluctuations mean value over the ensemble, infact 
\begin{equation}
\begin{split}
G_\beta(\gamma,h,\beta)&=\sum_{n=1}^d \frac{(\partial_{\beta}B_n)^2}{B_n}=\\
&=\sum_{n=1}^d B_n ( E_n^2 + (\partial_\beta \ln Z )^2 +  E_n \partial_\beta \ln Z )=\\
&= \overline{E^2} - \overline{E}^2 = \overline{\Delta E^2}
\label{EnFluct}
\end{split}
\end{equation}
In order to assess LMG chains with $N=2, 3, 4$ as quantum thermometers we 
evaluate the QFI and maximize its value by tuning the external field.
In Fig. \ref{fig:234termoptf} we show the optimal values of field as a 
function of the anisotropy  for different values of $\beta$ and for 
sizes of the LMG chain. In contrast to what happened for the estimation 
of the anisotropy, the optimal values of the field $h^*$ do not correspond to 
the critical ones. On the other hand, there is clear connection between 
the two concepts: for each {\em critical} line different {\em optimal} 
lines exists, correspondingly to slightly larger and slightly smaller 
values of the field. As the inverse temperature is increased, the optimal 
lines are smoothly deformed, approaching the corresponding critical one 
from above and below. This link between critical and optimal lines will 
be examined in more details later in this section. 
\par
The explicit expression of the QFI $G_\beta(\gamma,h^*,\beta)$ for $N=2$
is given by 
\begin{align}
G_\beta =& \frac12 \frac{\kappa_3}{\kappa_4}
\label{gbeta2}
\end{align}
where
\begin{align}
\kappa_3 =& e^{\frac12 \beta (v + r)} \Big[
\frac12 (v+r)^2 + 4(1 + 8h^2 + \gamma^2) e^{\frac12 \beta  (v + r)}
+ \notag \\
& \frac12 (v - r)^2 e^{\beta (v+r)}
+ \frac12 (v + r)^2 e^{\beta r}
+ \frac12 (v + r)^2 e^{\beta v} \Big] \notag\\
\kappa_4 =& \Big[
e^{\frac12 \beta v} + e^{\frac12 \beta r} + e^{\frac12 \beta (v+ 2 r)} +
e^{\beta (v + \frac12 r)}
\Big]^2 \notag\,,
\end{align}
with $v$ and $r$ as in Eq. (\ref{gopt2}).
Analogue expressions, with several more terms, are obtained for $N=3$
and $N=4$: we are not showing the explicit expressions here.
In the low temperature regime Eq. (\ref{gbeta2}) may be rewritten
as 
\begin{align}\label{gbasy}
 G_\beta \simeq \frac14 (v - r)^2\left\{
      \begin{array}{lr}
             e^{\frac12 \beta (v -  r)} &  h \geq \sqrt{\gamma}/2 \\
	            e^{-\frac12 \beta (v-r)}  & h < \sqrt{\gamma}/2
		         \end{array}
			    \right.\,,
			    \end{align}
where, as in Eq. (\ref{ggasy}),
the exponent is the energy gap between the two lowest energy levels.
\par
\begin{figure}[h!]
\includegraphics[width=0.48\columnwidth]{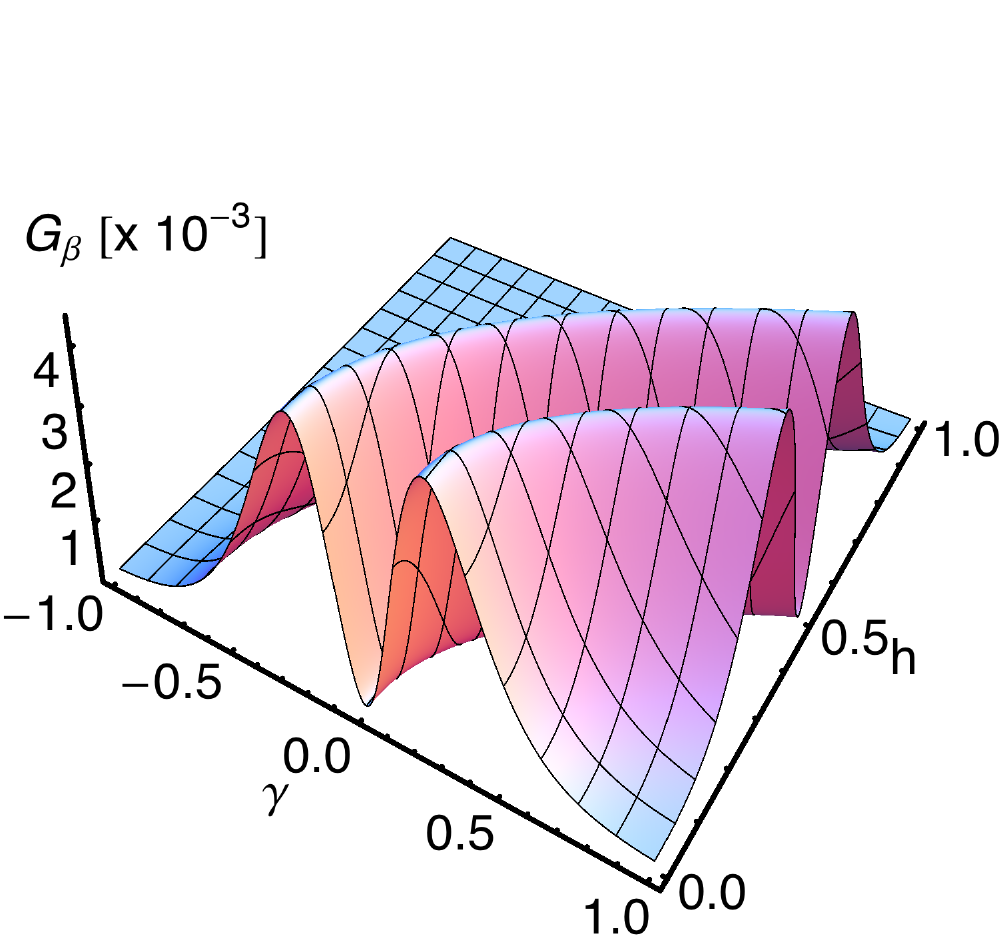}
\includegraphics[width=0.48\columnwidth]{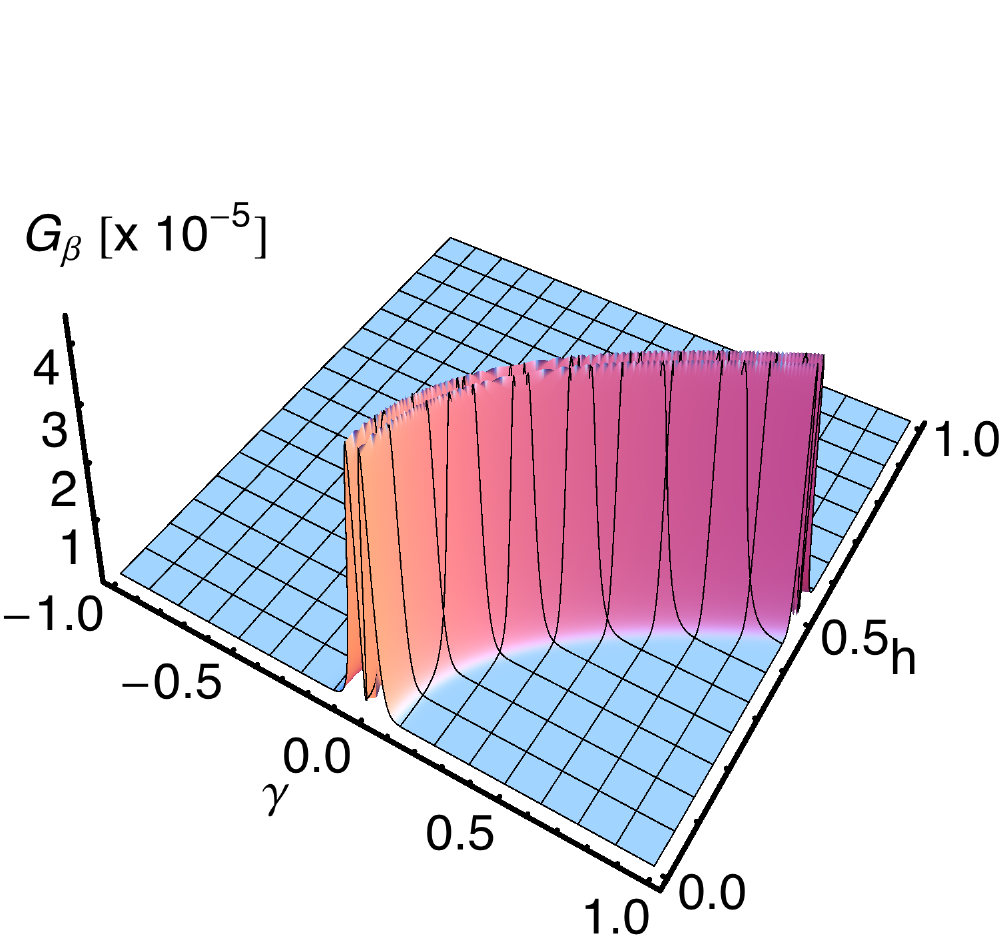}
\includegraphics[width=0.48\columnwidth]{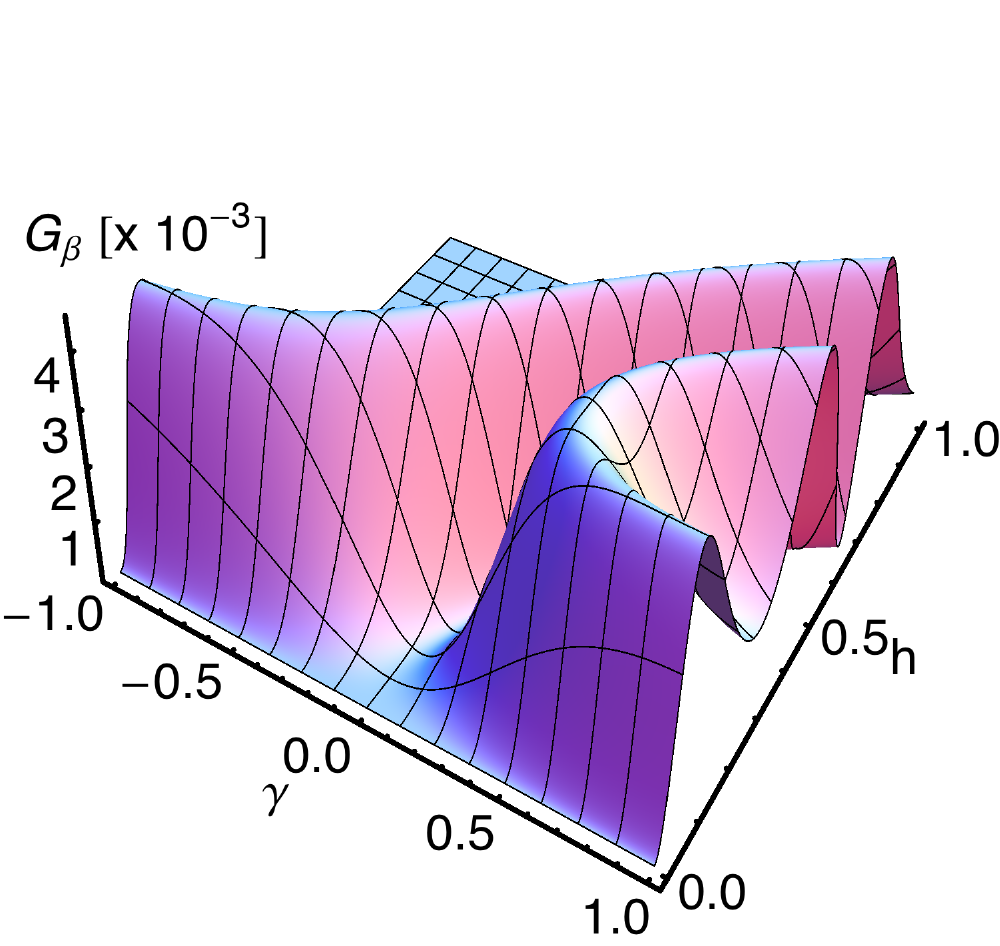}
\includegraphics[width=0.48\columnwidth]{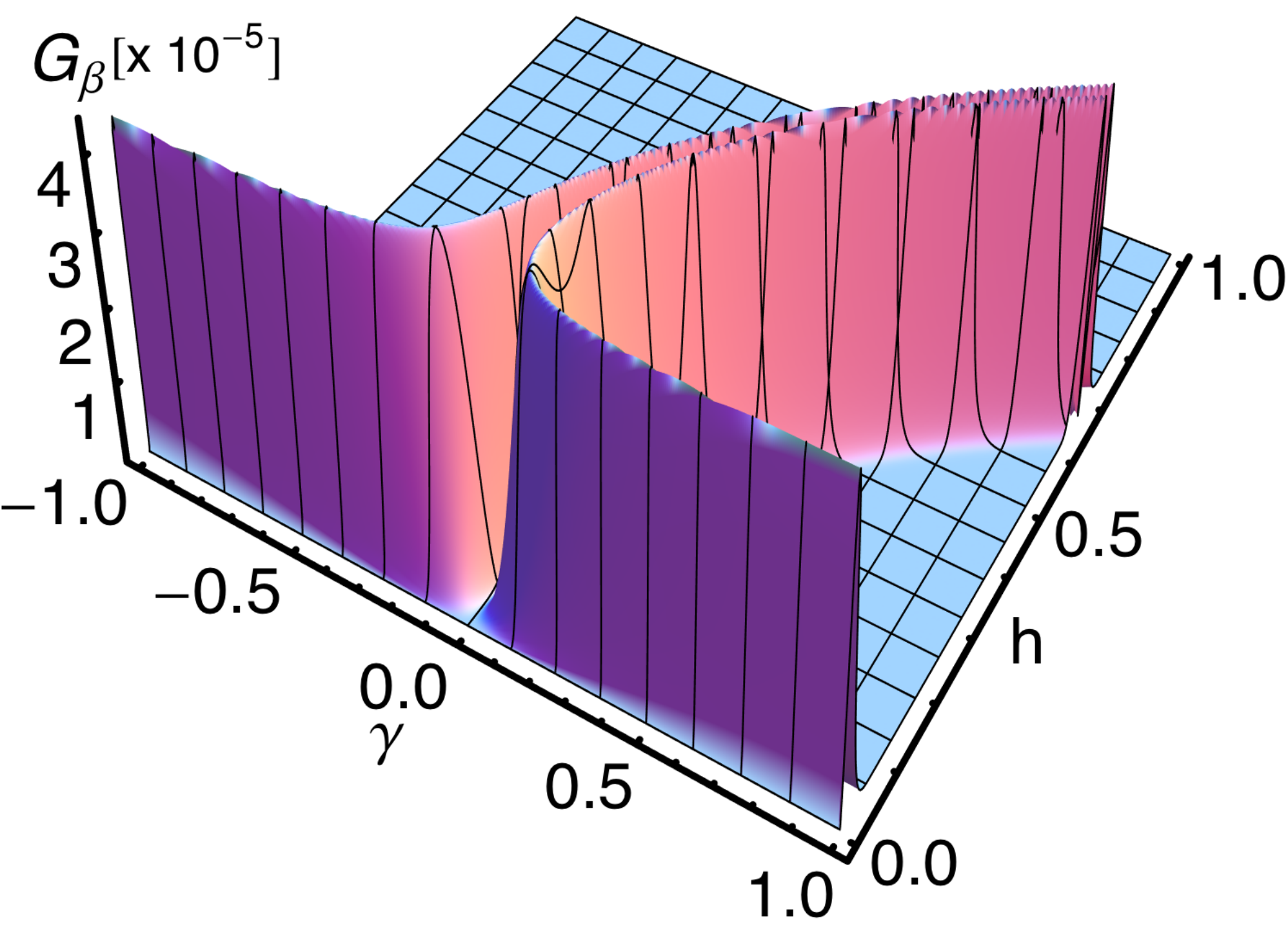}
\includegraphics[width=0.48\columnwidth]{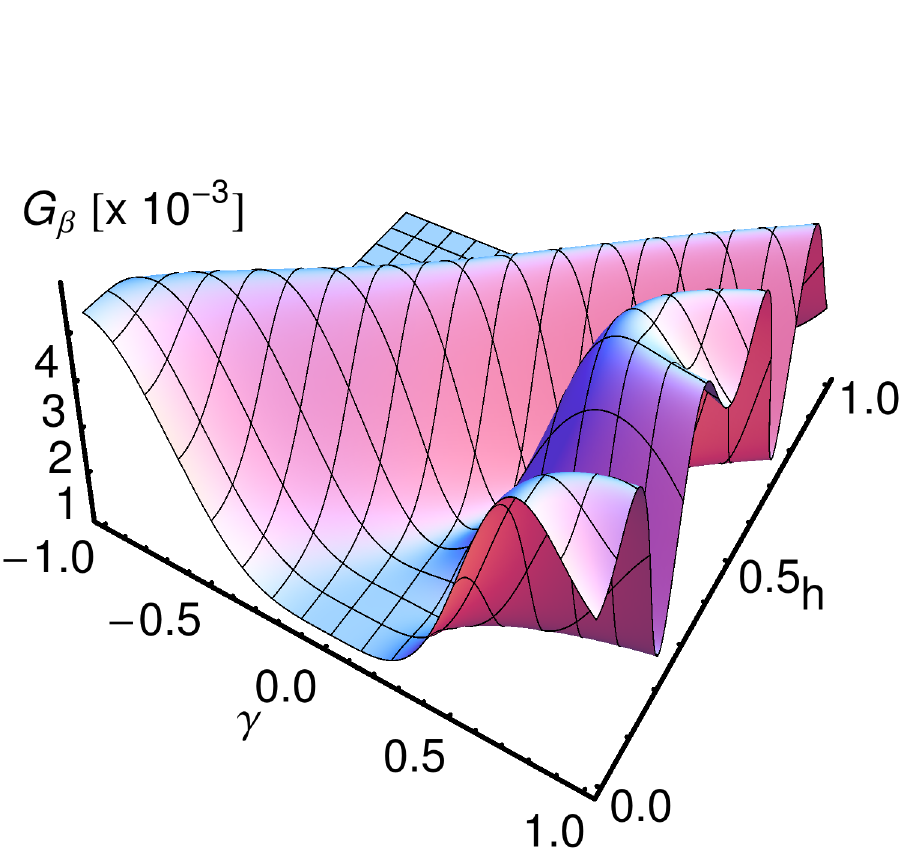}
\includegraphics[width=0.48\columnwidth]{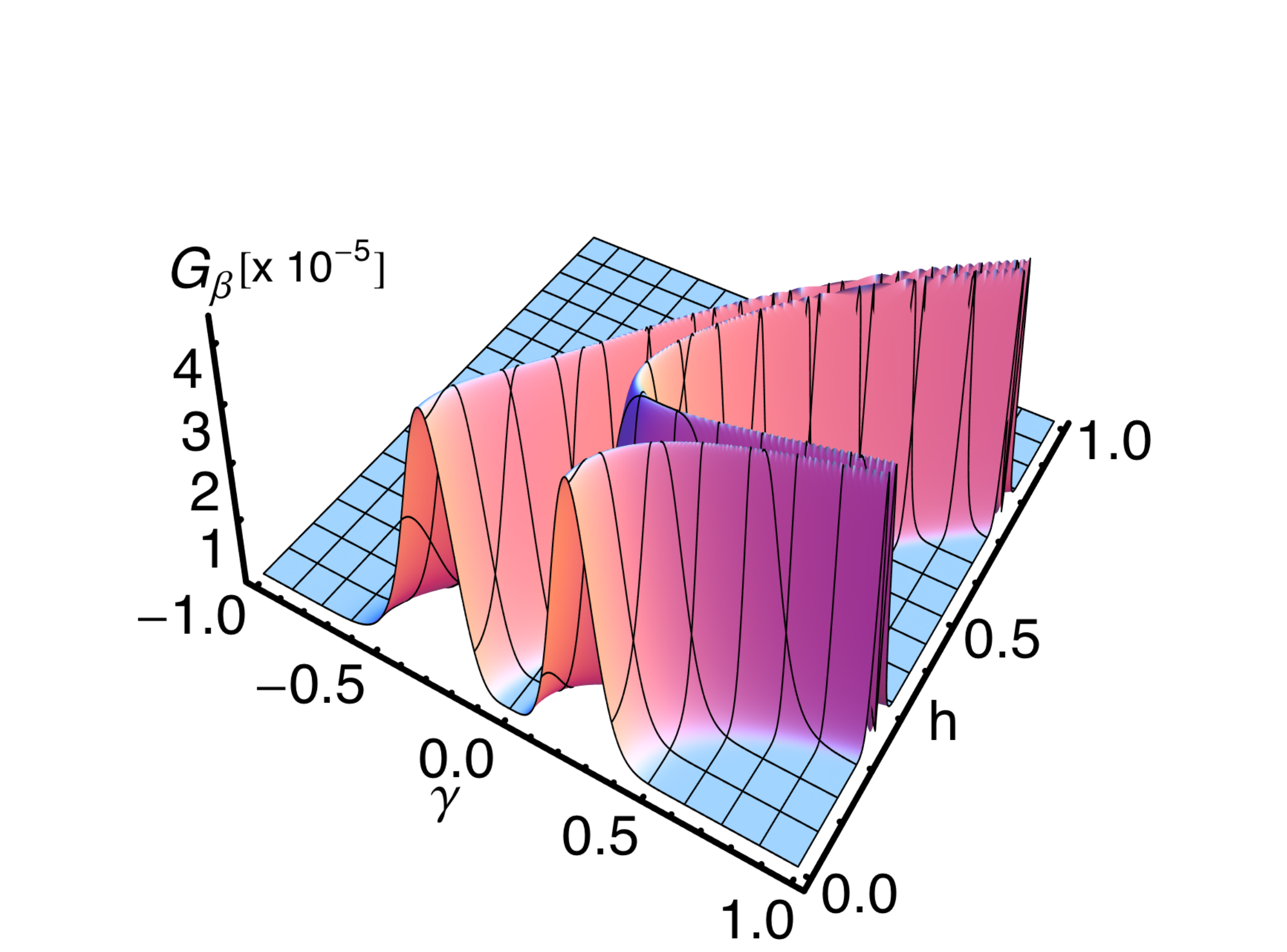}
\caption {Quantum thermometry using LMG systems. The plots show
$G_\beta$ versus $\gamma$ and $h$ for different $\beta$ and number of sites.
The three rows report results for $N=2,3,4$ respectively. The two columns 
refer to $\beta= 10$ and $\beta= 100$.}
  \label{fig:234term}
\end{figure}

In order to gain more insight on the QFI behaviour, in Fig.
\ref{fig:234term}  we show $G_\beta$  as a function of the anisotropy
and of the external field for different values of $\beta$ and the the
number of sites. At first we notice that the presence of optimal lines
clearly emerges from the plot. The QFI decreases with $\beta$ for any
value of the anisotropy and the external field and  this may easily
understood intuitively: as temperature decreases $\rho(\gamma,h,\beta)$
approaches the projector on the GS space and being this projector
independent on the temperature, the QFI vanishes.  On the other hand,
the quantitative features of the decay, e.g. how fast the optimal
$G_\beta$ tends to zero, are strongly influenced by the criticality of
the system. Indeed, outside the critical regions the QFI vanishes
exponentially, whereas along the optimal lines it vanishes as
$\frac{1}{\beta^2}$ independently on $\gamma$.  For increasing $\beta$
two phenomena occur: i) the optimal lines approach the critical 
ones, $h^*\rightarrow h_c$;
ii) the QFI $G_\beta$ shows a behavior independent on $N$, i.e. 
Eq. (\ref{gbasy}) may generalized to $N=3, 4$ and 
rewritten as
\begin{equation}
G_\beta \simeq k(\gamma,h) e^{-\beta \ f(\gamma,h) }
\end{equation}
where the functions $k(\gamma,h)$ and $f(\gamma,h)$ are non negative, 
independent on $\beta$ and zero only on the critical/optimal lines. 
Overall, we have that the presence of degeneracy, i.e. crossing between
the lowest eigenvalues, allow us to find optimal fields where $G_\beta$
decreases as $ 1/\beta^2$, suggesting that the criticality itself is 
the reason behind such enhancement. 
\par
In order to confirm this intuition and to gain more insight on the QFI
behavior in the low temperature regime we again consider  the two-level
approximation used before. Using the notation of Eq. (\ref{eq31}), the
QFI rewrites
\begin{equation} 
G_\beta(a,b,\beta)=\frac{e^{\beta x(a,b)} 
\left[\beta x(a,b)\right]^2 }{\left[1 + 
e^{\beta x(a,b)}\right]^2}  \frac{1}{\beta^2} 
= \frac{F(\beta x(a,b))}{\beta^2}\,,
\end{equation}
where $F(y)$ is a symmetric function vanishing in the origin, $F(0)= 0
$, and it shows two global maxima at $y=\pm y_{opt}$. This explains the
behavior shown in Fig.~\ref{fig:234termoptf} and~\ref{fig:234term} where
for each critical line, i.e. $x(a,b)=0$, two optimal lines
are present, corresponding to $\beta x(a,b)~=~\pm y_{opt}$.
Moreover, the dependence of $F(y)$ on the product of  $\beta$ with
$x(a,b)$ clarifies why, as $\beta$ increases, the optimal lines approach
the critical ones. Finally, we see that on the optimal lines the 
QFI vanishes as $1/\beta^2$ independently on any parameter, since 
the maximization of $F(y)$ factored out the parameter dependence. In 
other words, the precision is basically governed by the energy gap 
between the two lowest energy levels.  This behavior, in the limit of 
large $\beta$, is independent on the actual model, so that the argument 
may be equally employed to describe any system with an energy spectrum
made of two crossing lowest levels well separated from the other
levels.  
\par
We finally emphasize that the ultimate bound to precision may be
practically achieved, since, as shown by Eq. (\ref{EnFluct}) the SLD
turns out to be the total energy of the system, which we assume to be
measurable.
\section{Robustness against fluctuations of the external field}
\label{hfl}
The results reported in the previous Sections shows that criticality
is a resource for quantum metrology in LMG systems. As it has been
extensively discussed, in order to achieve the ultimate bounds to 
precision one should tune the external field to the 
appropriate value, driving the system towards the critical region.
A question thus arises on whether and how an imprecise tuning of the
external affects the metrological performances of the system.
\par
This issue basically amounts to a perturbation
analysis in order to discuss the robustness of the optimal estimators
against fluctuations of the external field. 
The canonical approach to attack this problem would be that of 
considering the state of the system as a mixture of different ground 
states, each one corresponding to a different value of the external 
field, and then evaluating the quantum Fisher
information for this family of states. This is a very challenging
procedure to pursue, even numerically, and some approximated approach
should be employed instead. In fact, it is possible 
to provide an estimate of this effect by averaging the QFI
over a given distribution for the external field: this is an
approximation since the QFI is a nonlinear function of the density
operator, but it is not a crude one, owing to the small value of
fluctuations that we should consider for this kind of perturbation
analysis. 
\par
\begin{figure}
\includegraphics[width=0.9\columnwidth]{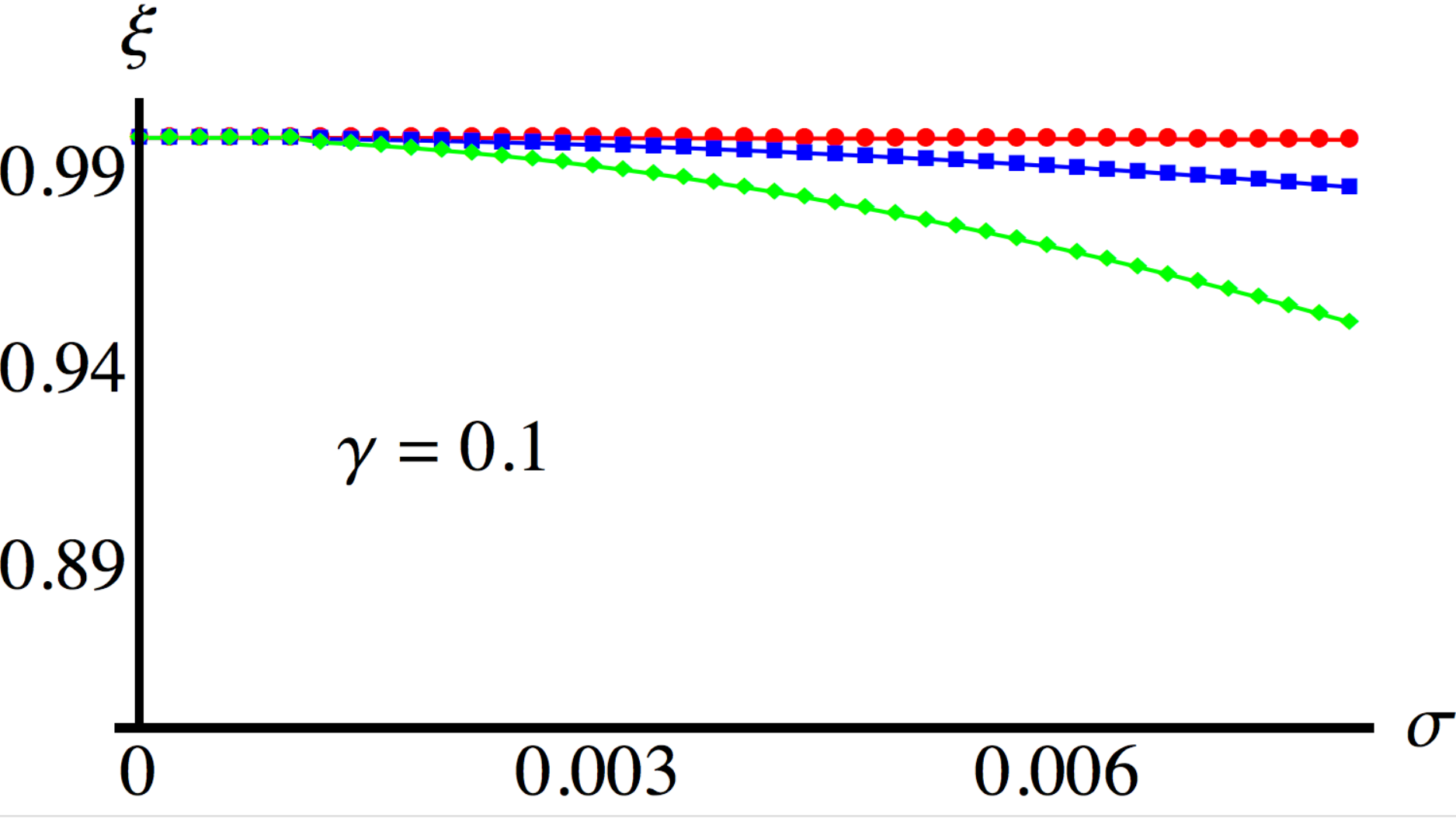}
\includegraphics[width=0.9\columnwidth]{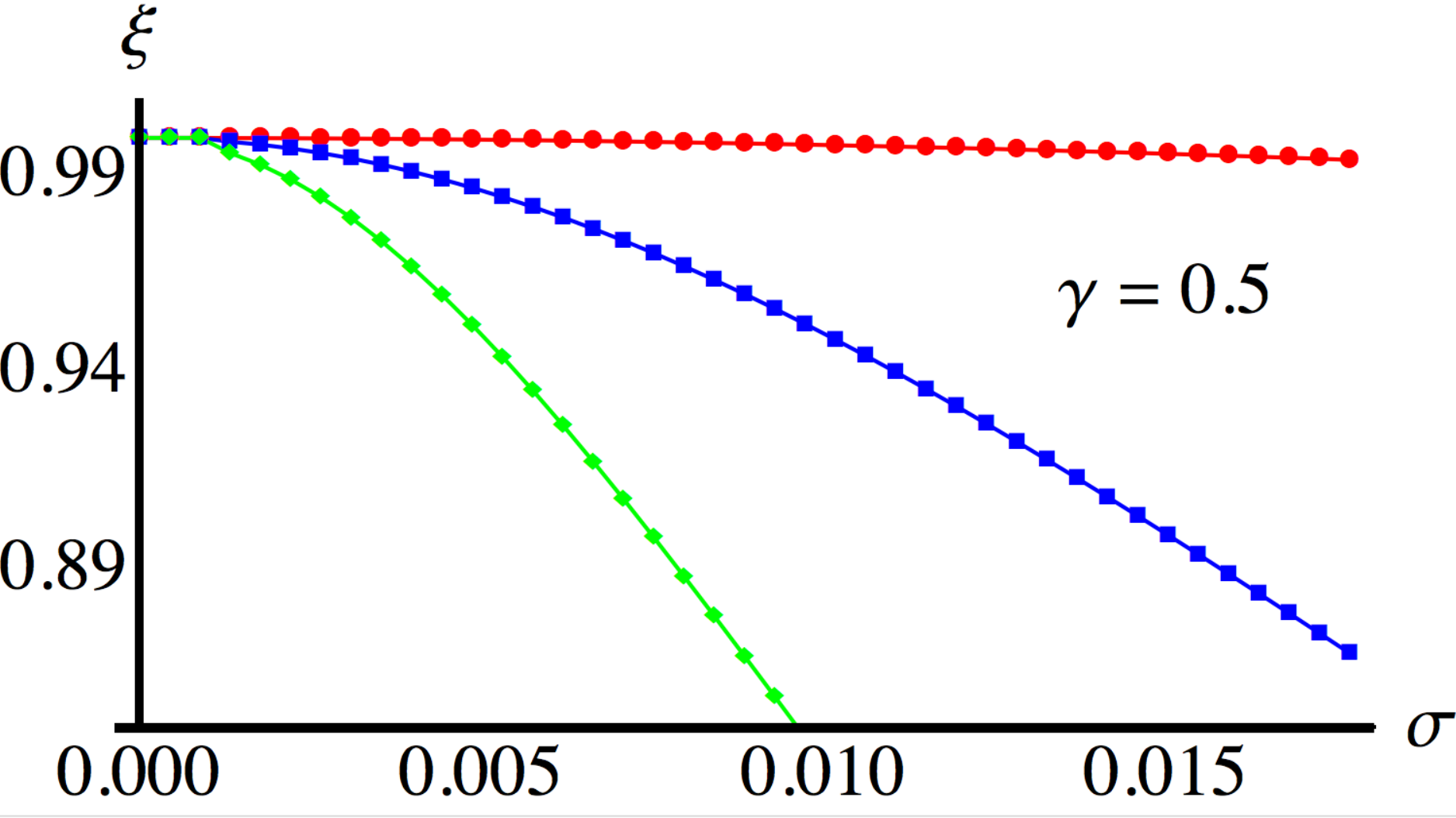}
\caption{
The ratio
$\xi = \xbar{G}_\gamma (\beta)/G_\gamma(\gamma,h_c,\beta)$
between the field-averaged QFI and the optimal one
as a function of the width $\Sigma$ of the field 
distribution. The upper panel show results for $\gamma=0.1$ and
the lower one for $\gamma=0.5$. In both panels we show the behavior
for $\beta=5$ (red points), $\beta=25$ (blue squares), and $\beta=50$ 
(green diamonds).
}\label{f:hfl}
\end{figure}

In order to obtain a quantitative estimate we assume that the 
actual value of the external field is normally distributed around 
the optimal one $h_c$, 
and evaluate the averaged QFI for the anistropy 
\begin{align}
\xbar{G}_\gamma(\beta) = \int\! dh\, G_\gamma (\gamma,h,\beta)\, g_\Sigma(h)
\end{align}
as a function of the width $\Sigma$ of the Gaussian $g_\Sigma (h)$, 
viewed as a convenient measure of the fluctuations (i.e. of the imprecise 
tuning) of the external
field. In particular, we choose the range of $\Sigma$ as to describe 
an imprecise tuning of the external field up to $\pm 5\%$.
In Fig. \ref{f:hfl} we show the ratio 
between the field-averaged QFI and the optimal one
\begin{align}
\xi = \frac{\xbar{G}_\gamma (\beta)}{G_\gamma(\gamma,h_c,\beta)}\,,
\end{align}
as a function of the of the width $\Sigma$ of the Gaussian distribution,
for different value of $\gamma$ and for different temperatures.
As it is apparent from the plots, the ratio is close to unit, showing
the robustness of the optimal estimator. The plots also show that the 
detrimental effects of an imprecise tuning of $h$ increase with $\gamma$
and decrease with temperature.
Analogue results may be obtained for $N=3$ and $N=4$ as well
as for the estimation of temperature.
Overall, we have that the optimal estimators are robust against
possible fluctuations of the external field, thus providing a realistic
benchmark for precision measurements on LMG systems.
\section{Quantum estimation in large LMG system: the thermodynamical limit}
\label{thermo}
The study of the thermodynamical limit of the model could be conducted
using the diagonal form of the Hamiltonian in Eq.(~\ref{eq:Hdiag}).  The
family of quantum states we are dealing with may be expressed as
$\rho_\Theta = U_\Theta \rho (\gamma,h,\beta) U_\Theta^\dagger$ where
$U_\Theta = \exp{(-i \Theta(\gamma,h) G )}$ is a unitary operator, $G
\equiv (a^2 + a^{\dagger 2})$ is the Hermitian operator related to the
Bogolyubov transformation in Eq.(~\ref{Bog}).  This let us to compute
the QFIs for anisotropy $G_\gamma$ and temperature $G_\beta$ using
Eq.(~\ref{QFIfinal}), where the parameter $\lambda$ turns out to be in
the first case $\gamma$ and in the second the inverse temperature
$\beta$. It is useful to underline that, in the limit of an infinite
number of particle the sum in Eq.(~\ref{QFIfinal}) is infinite thus
leading to region where the quantum Fisher information is divergent.
\par
\begin{figure}[h!] 
\includegraphics[width=0.48\columnwidth]{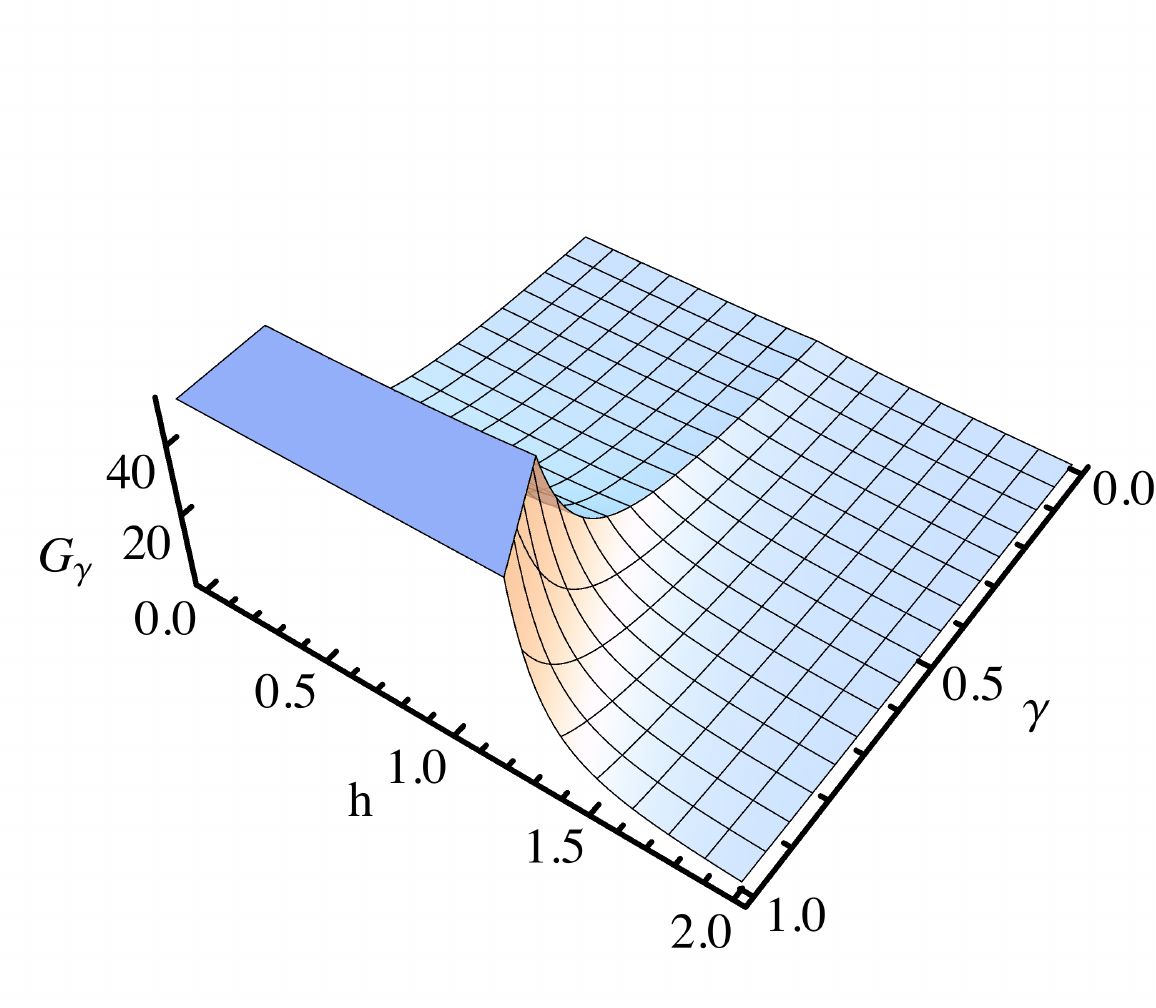}
\includegraphics[width=0.48\columnwidth]{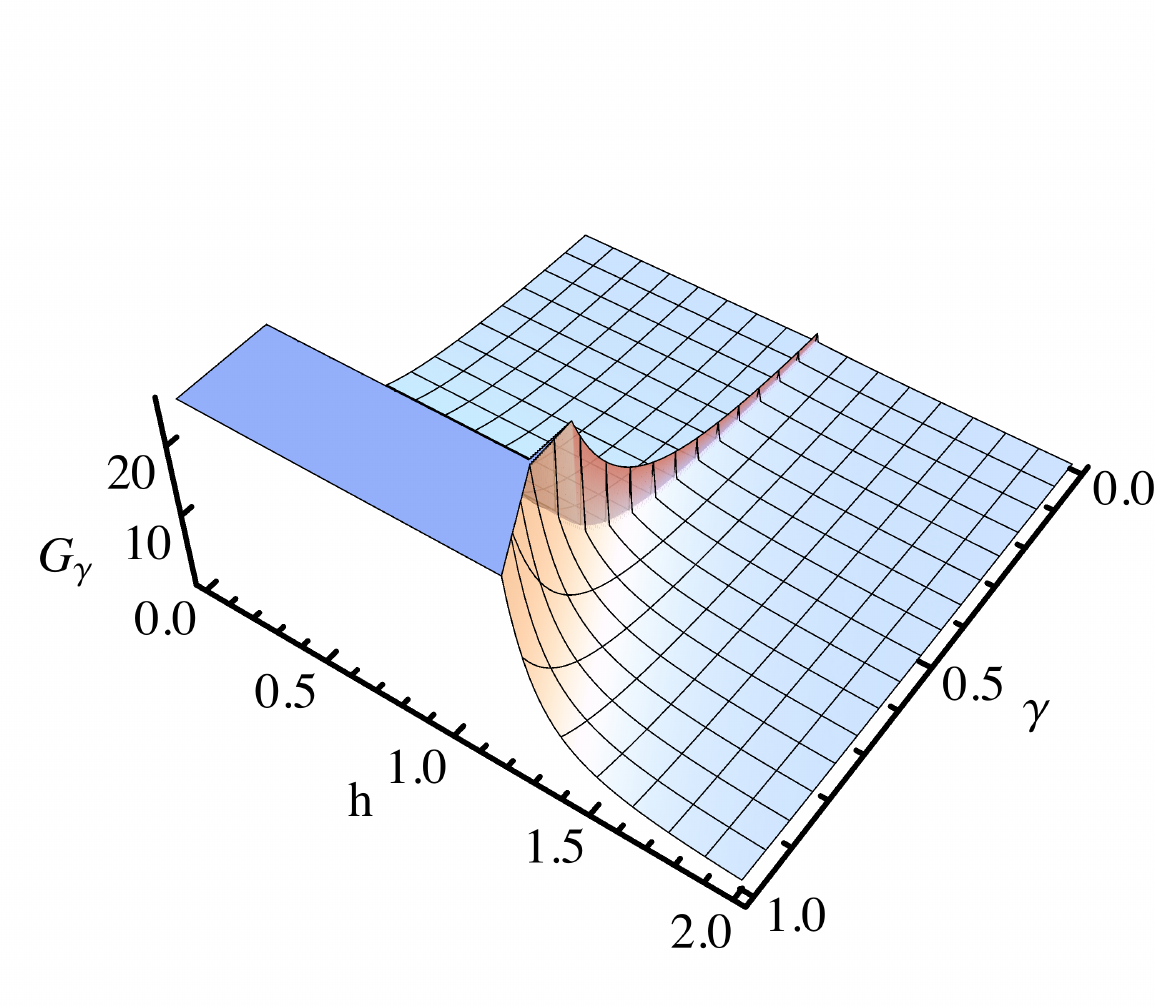}
\caption {Characterization of anisotropy in the thermodynamical limit.
The plots show the behavior of  $G_\gamma$ for the LMG model as function
of the anisotropy parameter $\gamma$ and the external magnetic field
$h$. The left panel refers to $\beta=1$ and the right one to
$\beta=10^5$.} \label{GgTh}
\end{figure}

We do not report here the analytic expressions of the QFIs since they are quite cumbersome. Rather we discuss their behavior analyzing their main features.
In  Fig.~\ref{GgTh} we show $G_\gamma$ as a function of the external field $h$ 
and of the anisotropy $\gamma$ itself. As it is apparent from the plot, 
in the ordered phase ($h>1$) $G_\gamma$ has a finite value everywhere, showing a cusp for $h$ approaching the critical value.
In the broken phase $G_\gamma$ increases with $\gamma$ showing a divergent behaviour approaching $\gamma=1$ for all value 
of the magnetic field in the region, thus signaling the sudden change of universality 
class of the system. In both phases the scaling with the temperature on the critical 
regions goes as $\beta^2$. More specifically, we 
have 
\begin{equation}
G_\gamma (\gamma,h^*,\beta) \simeq
 \frac{9}{4 (h - 1)^2} - \frac{25 \beta^2}{12} + O(h), 
\label{gammascaling1}
\end{equation}
in the orderd phase, $h>1$ and 
\begin{equation}
G_\gamma (\gamma,h^*,\beta) \simeq
 \frac{9}{4 (\gamma - 1)^2} - \frac{25 \beta^2(h - 1)}{6(\gamma - 1)} + O(h)\,,
\label{gammascaling2}
\end{equation}
in the broken one, i.e. for $0\leq h < 1$.
\par
The evaluation of the quantum Fisher information for the temperature
shows how it reaches is maximum, without showing  divergences, along the
degeneracy lines previously outlined, but this time it scales as
$\beta^{-2}$ at the first order near the critical field. If $ h \geq1$
we have \begin{equation}
G_\beta (\gamma,h,\beta) \simeq
 \frac{1}{\beta^2} + \frac{1}{3}(\gamma -1 )(h-1) + O(h^\frac{3}{2})
\label{Tscaling1}
\end{equation}
instead in the other phase where $0 \leq h <1$ we obtain 
\begin{equation}
G_\beta (\gamma,h,\beta) \simeq
 \frac{1}{\beta^2} - \frac{2}{3}(\gamma -1 )(h-1) + O(h^\frac{3}{2})\,.
\label{Tscaling2}
\end{equation}
We notice that this results could be improved only going beyond the Gaussian approximation performed in Eqs.(~\ref{HolsPrim},~\ref{Bog}) since in the broken 
phase region the effective separation between the degenerate ground
state vanishes as $\exp(-N)$. As a matter of fact, it would be possible
to recover the results obtained for the finite chain cases, i.e.
divergences along $h^* \simeq \sqrt{\gamma}$, only 
looking at the fine structure of the level in the broken phase.
\section{Conclusions}
\label{conclusion}
We have addressed quantum metrology in LMG model as a paradigmatic example 
of criticality-assisted estimation in systems with interaction beyond the 
first-neighbor approximation. In particular, we analyzed in details the 
use of criticality in improving precision of measurement procedures aimed 
at estimating the anisotropy of the system or its temperature.
\par
Upon considering LMG systems in thermal equilibrium with the environment 
we have evaluated exactly the quantum Fisher information of small-size 
LMG chains made of $N=2, 3$ and $4$ lattice sites and analyzed the same 
quantity in the thermodynamical limit by means of a zero-th order 
approximation of the system Hamiltonian. In this way we proved that 
quantum criticality of the system represents a resource in estimating  
the anisotropy. In fact, the quantum Fisher information $G_\gamma$
is maximized at the critical lines,  where, in the low temperature
regime, it  diverges as  $\beta^2$, while being finite everywhere else.
We have then shown that the ultimate bounds to precision may be achieved 
by tuning the external field and by measuring the total magnetization 
of the system. 
\par
We have also addressed the use of LMG systems as quantum thermometers
showing that: i) precision is governed by the gap between the lowest
energy levels of the systems, ii) field-dependent level crossing
provides a resource to extend the operating range of the quantum
thermometer.  Our results are encouraging for the emergent field of
quantum thermometry. Indeed, despite the fact that the QFI $G_\beta$
vanishes everywhere for decreasing temperature, criticality continues to
represent resource: the QFI is maximized along optimal lines approaching
the critical ones for decreasing temperature, and there the optimal QFI
vanishes as  $1/\beta^2$ instead of exponentially.  
\par
Finally, we have introduced a simple model, based on a two-level
approximation of the system, which allows us to provide an intuitive
understanding of our findings for both $G_\gamma$ and $G_\beta$. Our
model also suggests that similar behaviors may be expected for a larger
class of critical systems with interaction beyond the first-neighbor
approximation. 
\section*{Acknowledgments}
We acknowledge A. Lascialfari and G. Col\'o for useful discussions. 
This work has been supported by 
the MIUR project FIRB-LiCHIS-RBFR10YQ3H. 
\appendix
\section{LMG systems with $N=2, 3,  4$ sites}
\label{a:DD}
Here we provide the explicit expression, in the computational basis, 
of the Hamiltonian for LMG systems with $N=2, 3, 4 $ sites, as well as 
the eigenvalues and eigenvectors for $N=2, 3$. 
Throughout the Section we use the shorthand $u = (\gamma - 1)$ and $v = (\gamma + 1)$.
\subsection{N=2}
The matrix form of the two-site LMG Hamiltonian in the computational 
basis reads as follows
\begin{equation} H_2=-\frac12
\begin{pmatrix} 
  4h & 0 & 0 & u\\
  0&0&v&0 \\
  0 & v & 0  & 0\\
u & 0 & 0 & -4h
 \end{pmatrix}\,.
\end{equation} 
The eigenvalues are given by
\begin{align}
\lambda_1 &= - \frac12 v  \quad \lambda_3 = - \frac12 \sqrt{16 h^2 + u^2} \\
\lambda_2 &=  \frac12 v \quad 
\lambda_4 =   \frac12 \sqrt{16 h^2 + u^2} \,,
\end{align}
and the corresponding (unnormalized) eigenvectors by
\begin{align}
\mathbf{u_1^T} &= \begin{pmatrix}   0, & 1, & 1, & 0 \end{pmatrix}  \\
\mathbf{u_2^T} &=  \begin{pmatrix}   0, & -1, & 1, & 0 \end{pmatrix}  \\
\mathbf{u_3^T}&= \begin{pmatrix}   \frac{4h + \sqrt{16 h^2 + u^2}}{u}, & 0, & 0, & 1 \end{pmatrix} \\
\mathbf{u_4^T}&=   \begin{pmatrix}   \frac{4h - \sqrt{16 h^2 + u^2}}{u}, & 0, & 0, & 1 \end{pmatrix}\,.
\end{align}
\subsection{N=3}
The Hamiltonian for the three-site LMG system is given by
\begin{equation} H_3= -\frac{1}{3}
\begin{pmatrix} 
  9h & 0 & 0 & -u& 0  & -u& -u& 0 \\
  0 &  3h  & v& 0  & v&0&0 & -u \\
    0 & v&3h & 0  &  v & 0 & 0 & -u\\
  - u & 0 & 0 & -3h&0& v& v  & 0 \\
    0 & v & v& 0 & 3h& 0  & 0& -u \\
 -u & 0 & 0 & v &0& - 3h & v  & 0 \\
-u & 0 & 0 & v& 0  & v&-3h&0 \\
0  & -u & -u & 0  &- u& 0&0  & -9h\\
 \end{pmatrix}\,,
\end{equation}
leading to the eigenvalues 
\begin{align}
&\mu_{1,2} =  \frac13 (v-3h)  \quad
\mu_{3,4} =  \frac13 (v+3h) \\
&\mu_5 =  \frac13 (- 3h - v -\Delta_- )\\
&\mu_6 =  \frac13 (- 3h - v + \Delta_- )\\
&\mu_7 =  \frac13 (3h - v  -\Delta_+)\\
&\mu_8 =  \frac13 (3h - v  +\Delta_+)
\end{align}
and eigenvectors
\begin{align}
\mathbf{v_1^T} &= \left(0,-1,0, 0, 1, 0, 0, 0 \right)  \\
\mathbf{v_2^T} &=  \left(  0, -1, 1, 0, 0, 0, 0, 0 \right)  \\
\mathbf{v_3^T}&= \left( 0, 0, 0, -1, 0, 0, 1, 0 \right)  \\
\mathbf{v_4^T}&=   \left( 0, 0, 0, -1, 0, 1, 0, 0 \right) \\
\mathbf{v_5^T}&=   \begin{pmatrix}   \frac{\delta_+-\Delta_-  }{u},& 0,& 0,& 1,& 0,& 1,& 1,& 0 \end{pmatrix}\\
\mathbf{v_6^T}&=   \begin{pmatrix}   \frac{\delta_+ +\Delta_- }{u},& 0,& 0,& 1,& 0,& 1,& 1,& 0 \end{pmatrix}\\
\mathbf{v_7^T}&=   \begin{pmatrix} 0,&  \frac{\delta_-- \Delta_+}{3u},& \frac{\delta_-- \Delta_+}{3u},& 0,& \frac{ \delta_-- \Delta_+}{3u},& 0,& 0,& 1 \end{pmatrix}\\
\mathbf{v_8^T}&=   \begin{pmatrix} 0,& \frac{\delta_-+ \Delta_+}{3u},& \frac{\delta_-+ \Delta_+}{3u},& 0,& \frac{\delta_-+ \Delta_+}{3u},& 0,& 0,& 1 \end{pmatrix}
\end{align}
where $\Delta_\pm = 2 \sqrt{1+9h^2 \pm 3 h v+\gamma u }$ and 
$\delta_\pm = - 6h \pm v$.
\subsection{N=4}
The Hamiltonian of a four-site LMG system may be expressed in a block-diagonal 
form given by
\begin{equation}  H_4= 
\begin{pmatrix} 
  A & 0 & \cdots  & 0 \\
  0  & B & \cdots & 0\\
  0& \cdots & B  & 0\\
  0 & \cdots &   0 & C
 \end{pmatrix}
\end{equation}
where 
\begin{equation} A = -\frac{1}{4}
\begin{pmatrix} 
  16 h & 0 & -\sqrt{6} u &0& 0\\
  0& 3v+ 8h &0&-3u&0\\
  -\sqrt{6}u & 0 & 4v & 0 &-\sqrt{6}u\\
    0& -3u&0&   3 v - 8h &0\\
      0 & 0 & -\sqrt{6} u &0& -16 h\\
   \end{pmatrix}
\end{equation}

\begin{equation} B = \frac{1}{4}
\begin{pmatrix} 
v - 8h & 0 & u\\
  0&0&0\\
  u& 0& v + 8h
   \end{pmatrix}
\end{equation}

\begin{equation} C = \frac{1}{4}
\begin{pmatrix} 
2v&0&0&0&0\\ 
 0 & v- 8h & 0 & u & 0\\
 0&0& 0&0&0\\
 0& u & 0&  v+8h &0\\
0&0&0&0&  2v \\
   \end{pmatrix}.
\end{equation}
\par

\end{document}